\documentclass[epj]{svjour}
\usepackage{graphics}
\usepackage{amsmath}
\usepackage{amssymb}
\begin{document}
\title{Momentum dependence of the imaginary part of the $\omega$- and $\eta^\prime$-nucleus optical potential}

\author{
S.~Friedrich$^{1}$,~M.~Nanova$^{1}$,~V.~Metag$^{1}$,~F.~N.~Afzal$^{2}$,~D.~Bayadilov$^{2,3}$,~B.~Bantes$^{4}$,~R.~Beck$^{2}$,~M.~Becker$^{2}$,~S.~B\"ose$^{2}$,\\K.-T.~Brinkmann$^{1}$,~V.~Crede$^{5}$,~P.~Drexler$^{1,a}$,~H.~Eberhardt$^{4}$,~D.~Elsner$^{4}$,~F.~Frommberger$^{4}$,~Ch.~Funke$^{2}$,~M.~Gottschall$^{2}$,\\M.~Gr\"uner$^{2}$,~E.~Gutz$^{1}$,~Ch.~Hammann$^{2}$,~J.~Hannappel$^{4}$,~J.~Hartmann$^{2}$,~W.~Hillert$^{4}$,~Ph.~Hoffmeister$^{2}$,~Ch.~Honisch$^{2}$,\\T.~Jude$^{4}$,~D.~Kaiser$^{2}$,~F.~Kalischewski$^{2}$,~I.~Keshelashvili$^{6,b}$,~F.~Klein$^{4}$,~K.~Koop$^{2}$,~B.~Krusche$^{6}$,~M.~Lang$^{2}$,~K.~Makonyi$^{1,c}$,\\F.~Messi$^{4}$,~J.~M\"uller$^{2}$,~J.~M\"ullers$^{2}$,~D.-M.~Piontek$^{2}$,~T.~Rostomyan$^{6}$,~D.~Schaab$^{2}$,~Ch.~Schmidt$^{2}$,~H.~Schmieden$^{4}$,\\R.~Schmitz$^{2}$,~T.~Seifen$^{2}$,~V.~Sokhoyan$^{2,a}$,~C.~Sowa$^{7}$,~K.~Spieker$^{2}$,~A.~Thiel$^{2}$,~U.~Thoma$^{2}$,~T.~Triffterer$^{7}$,~M.~Urban$^{2}$,\\H.~van~Pee$^{2}$,~D.~Walther$^{2}$,~Ch.~Wendel$^{2}$,~D.~Werthm\"uller$^{6,d}$,~U.~Wiedner$^{7}$,~A.~Wilson$^{2}$,~L.~Witthauer$^{6}$,~Y.~Wunderlich$^{2}$, and~H.-G.~Zaunick$^{1}$\\
(The CBELSA/TAPS Collaboration)
\mail{Mariana.Nanova@exp2.physik.uni-giessen.de}}
\titlerunning{Momentum dependence of the imaginary part of the $\omega$- and $\eta^\prime$-nucleus optical potential}
\authorrunning{S.Friedrich \textit{et al.}}

\institute{
{$^{1}$II. Physikalisches Institut, Universit\"at Gie{\ss}en, Germany}\\
{$^{2}$Helmholtz-Institut f\"ur Strahlen- und Kernphysik, Universit\"at Bonn, Germany }\\
{$^{3}$Petersburg Nuclear Physics Institute, Gatchina, Russia}\\
{$^{4}$Physikalisches Institut, Universit\"at Bonn, Germany}\\
{$^{5}$Department of Physics, Florida State University, Tallahassee, FL, USA}\\
{$^{6}$Departement Physik, Universit\"at Basel, Switzerland}\\
{$^{7}$Physikalisches Institut, Universit\"at Bochum, Germany}\\
{$^{a}$Current address: Institut f\"ur Kernphysik, Universit\"at Mainz}\\
{$^{b}$Current address: Institut f\"ur Kernphysik, Forschungszentrum J\"ulich, Germany}\\
{$^{c}$Current address: Stockholm University, Stockholm, Sweden}\\
{$^{d}$Current address: School of Physics and Astronomy, University of Glasgow, UK}\\
}

\date{Received: date / Revised version: date}
%
\abstract{The photoproduction of $\omega$ and $\eta^\prime$ mesons off carbon and niobium nuclei has been measured as a function of the meson momentum for incident photon energies of 1.2-2.9 GeV at the electron accelerator ELSA. The mesons have been identified via the $\omega \rightarrow \pi^0 \gamma \rightarrow 3 \gamma$ and $\eta^\prime\rightarrow \pi^0 \pi^0\eta \rightarrow 6 \gamma$ decays, respectively, registered with the CBELSA/TAPS detector system. From the measured meson momentum distributions the momentum dependence of the transparency ratio has been determined for both mesons. Within a Glauber analysis the in-medium $\omega$ and $\eta^\prime$ widths and the corresponding absorption cross sections have been deduced as a function of the meson momentum. The results are compared to recent theoretical predictions for the in-medium $\omega$ width and $\eta^\prime$-N absorption cross sections. The energy dependence of the imaginary part of the $\omega$- and $\eta^\prime$-nucleus optical potential has been extracted. The finer binning of the present data compared to the existing data allows a more reliable extrapolation towards the production threshold. The modulus of the imaginary part of the $\eta^\prime$ nucleus potential is found to be about three times smaller than recently determined values of the real part of the $\eta^\prime$-nucleus potential, which makes the $\eta^\prime$ meson a suitable candidate for the search for meson-nucleus bound states. For the $\omega$ meson, the modulus of the imaginary part near threshold is comparable to the modulus of the real part of the potential. As a consequence, only broad structures can be expected which makes the observation of $\omega$ mesic states very difficult experimentally.}

%
\maketitle
\section{Introduction}
\label{intro}

The interaction of light pseudo-scalar and vector mesons with nucleons and nuclei and the possible existence of meson-nucleon clusters has recently been studied extensively experi$\-$mentally as well as theoretically \cite{HH,LMM,Oset,Zaki,Strakovsky,Moskal,Yamazaki_Akaishi,FINUDA,DISTO,Ichikawa,HADES,Fabbietti}. These investigations are motivated by the quest for the existence of mesic states, i.e. meson-nucleus bound states. The existence of deeply bound pionic states is clearly established \cite{Gilg,Itahashi,Geissel,Suzuki}. These systems are bound by the attractive Coulomb interaction between a negatively charged pion and the positively charged nucleus. The superposition with the strong interaction, which is repulsive at low pion momenta, leads to a potential pocket near the nuclear surface and consequently to a halo-like $\pi^-$ distribution
\cite{Kienle_Yamazaki}. Thus pions are only weakly absorbed, giving rise to rather narrow bound states which facilitated their experimental observation. 

The interaction of neutral mesons with nuclei has been studied to find out whether meson-nucleus states, only bound by the strong interaction, might exist as well. Here, one has to investigate whether the meson-nucleus interaction is sufficiently attractive and whether the meson absorption in nuclei is relatively weak to allow the formation of relatively narrow states. The interaction of mesons with nuclei can be described by an optical potential
\begin{equation}
U(r) = V(r) + i W(r), 
\end{equation}
comprising a real and an imaginary part, where $r$ is the distance of the meson to the centre of the nucleus. The depth of the real part $V(r)$  of the potential is a measure for the attraction and the size of the imaginary part $W(r)$ describes the strength of meson absorption. A necessary condition for the experimental observation of meson-nucleus bound states is that $\vert V \vert >> \vert W \vert $. It is therefore important to study experimentally the relative strength of the real and imaginary part of the meson-nucleus interaction for the meson of interest. It has been shown in \cite{Weil,Kotulla,Kotulla_err,Paryev,Nanova_realC,Nanova_tr,Metag_PPNP,Metag_Hypint} that the real part of the meson-nucleus potential can be extracted from measurements of the excitation function and momentum distribution of the meson while the imaginary part can be deduced from measurements of the transparency ratio which compares the meson production cross section off a nucleus with the one off the free nucleon.

When mesons are produced in a nuclear reaction in the 1-2~GeV energy range they exhibit - due to kinematics - a broad momentum distribution with a most probable value close to their mass. For the existence or non-existence of bound states, however, the potential parameters near threshold are decisive. Therefore, one has to measure the potential parameters over a wide range of meson momenta. It was the motivation for this work to improve earlier measurements of the momentum dependence of the imaginary part of the $\omega$- and $\eta^\prime$-nucleus potential \cite{Kotulla,Kotulla_err,Nanova_tr} by extending the momentum range, thereby facilitating a more reliable extra$\-$polation to low meson momenta. An extension to higher meson momenta is important for a dispersion relation analysis of the data which relates the imaginary part to the real part of the potential up to a constant, describing the non-dispersive meson-nucleus interaction. This approach will allow a consistency check for the determination of the real and imaginary part of the $\omega$- and $\eta^\prime$-nucleus potential obtained in independent measurements. The real part of the $\omega$ nucleus potential has already been studied at momenta as low as $\approx$ 300~MeV/$c$ \cite{Friedrich}. 

A further motivation for the present work were recent theoretical studies of the $\omega$ width in cold nuclear matter as a function of momentum and nuclear density by Ca$\-$brera and Rapp \cite{Cabrera_Rapp} and by Ramos {\it et al.} \cite{Ramos}.  Both groups calculated the in-medium $\omega$ width in a hadronic many-body approach, focusing on a detailed treatment of the in-medium modifications of intermediate $\pi \rho$ states. At normal nuclear matter density, they found an in-medium $\omega$ width of the order of 100-200~MeV, however, with differences in the 3-momentum dependence. The calculations were confined to the momentum ranges below 1000~MeV/$c$ in \cite{Cabrera_Rapp} and 600~MeV/$c$ in \cite{Ramos}. Earlier calculations investigated the direct coupling of the $\omega$ meson to nucleon resonances. Klingl et al. \cite{Klingl} and Lutz et al. \cite{Lutz} determined the $\omega$ nucleon scattering length and obtained both a much smaller value for the $\omega$ width of 40 MeV for vanishing $\omega$ momentum at normal nuclear matter density. M\"uhlich et al. \cite{Muehlich} calculated the momentum dependence of the $\omega$ width in a coupled-channel resonance model up to $\omega$ momenta of 600 MeV/c, starting from an in-medium width of 60 MeV for an $\omega$ meson at rest in the nucleus. In this experiment, the measurements extend to momenta of 2500~MeV/$c$, but particular attention has been paid to obtain reasonable statistics also in the low momentum range. For the $\eta^\prime$ meson, Oset and Ramos  \cite{Oset_Ramos} calculated in-medium $\eta^\prime$-nucleon inelastic cross sections, which are related to the in-medium $\eta^\prime$ width via the low density approximation. They found a weak $\eta^\prime$-nucleon coupling, leading to only small inelastic cross sections of the order of 3-15~mb. These theoretical predictions will be compared with the results of this work.

The paper is structured as follows: The experimental setup and the conditions of the experiment are described in section 2. Details of the analysis are given in section 3. Section 4 presents the experimental results and the comparison with the mentioned theoretical calculations. Concluding remarks are given in section 5.

\section{Experimental setup}
\label{expsetup}
The experiment was performed at the electron stretcher accelerator ELSA in Bonn \cite{Husmann_Schwille,Hillert}, using two solid state targets: carbon and niobium. Photons were produced, respectively, by scattering 3.2 and 3.0~GeV electron beams in a 50 $\mu$m thick copper radiator and in a 500 $\mu$m thick diamond radiator. The photons irradiated a carbon target of 15 mm thickness (5.9$\%$ of a radiation length $X_0$) and a niobium target of 1 mm thickness (8.6$\%$ of $X_{0}$), respectively. The bremsstrahlung photons were tagged in the energy ranges of 0.7-3.11~GeV for the carbon run and 1.2-2.9~GeV for the niobium run. Decay photons from mesons produced by the interaction in the target were detected with the combined Crystal Barrel (CB) (1320 CsI(Tl) modules) \cite{Aker} and MiniTAPS detectors (216 BaF$_2$ modules) \cite{TAPS1,TAPS2}. This detector setup subtended polar angles of 11$^{\circ}$-156$^{\circ}$ and 1$^{\circ}$-11$^{\circ}$, respectively, and the full azimuthal angular range, thereby \begin{figure*}
\begin{center}
 \resizebox{1.0\textwidth}{!}
 { \includegraphics{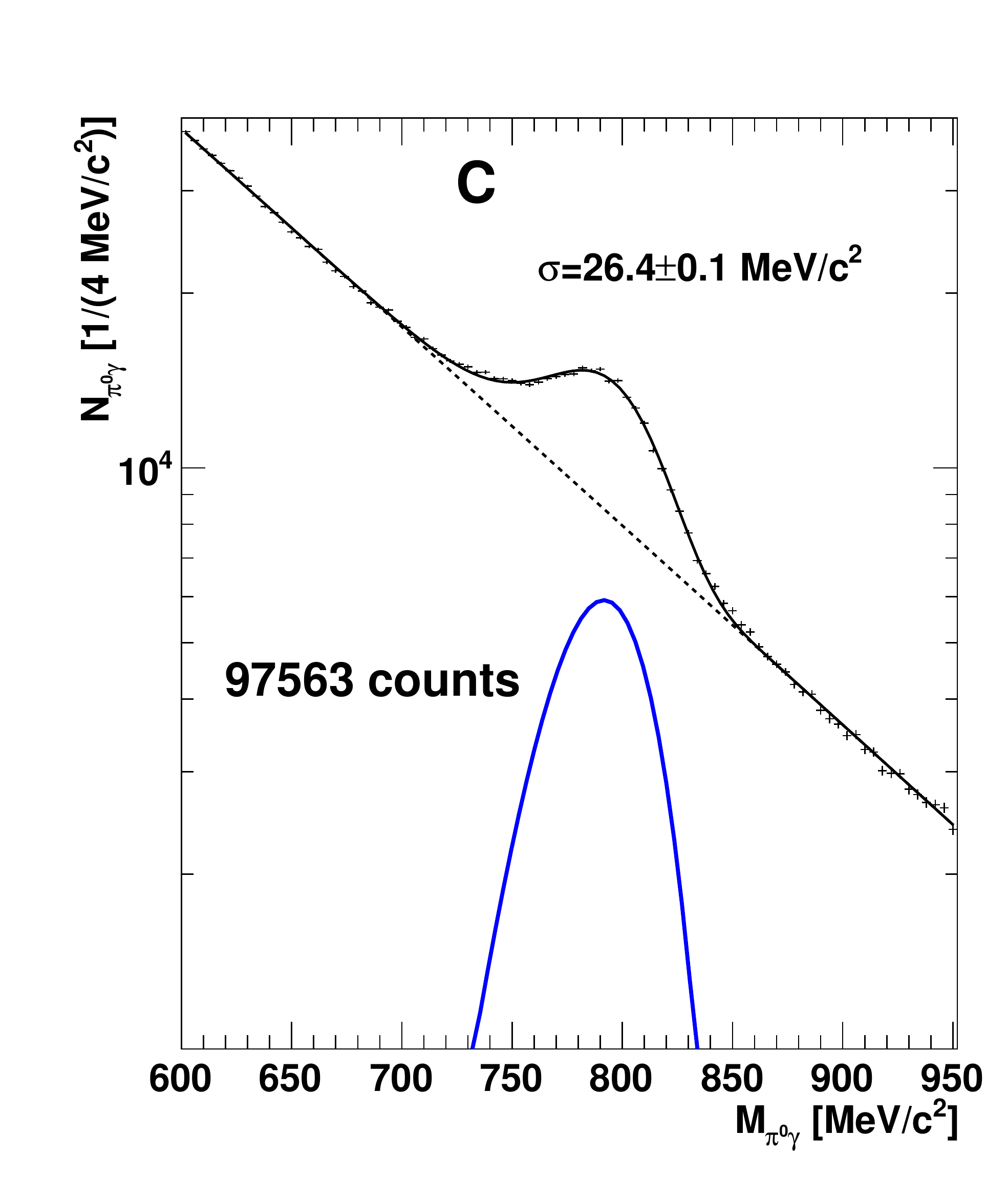} 
   \includegraphics{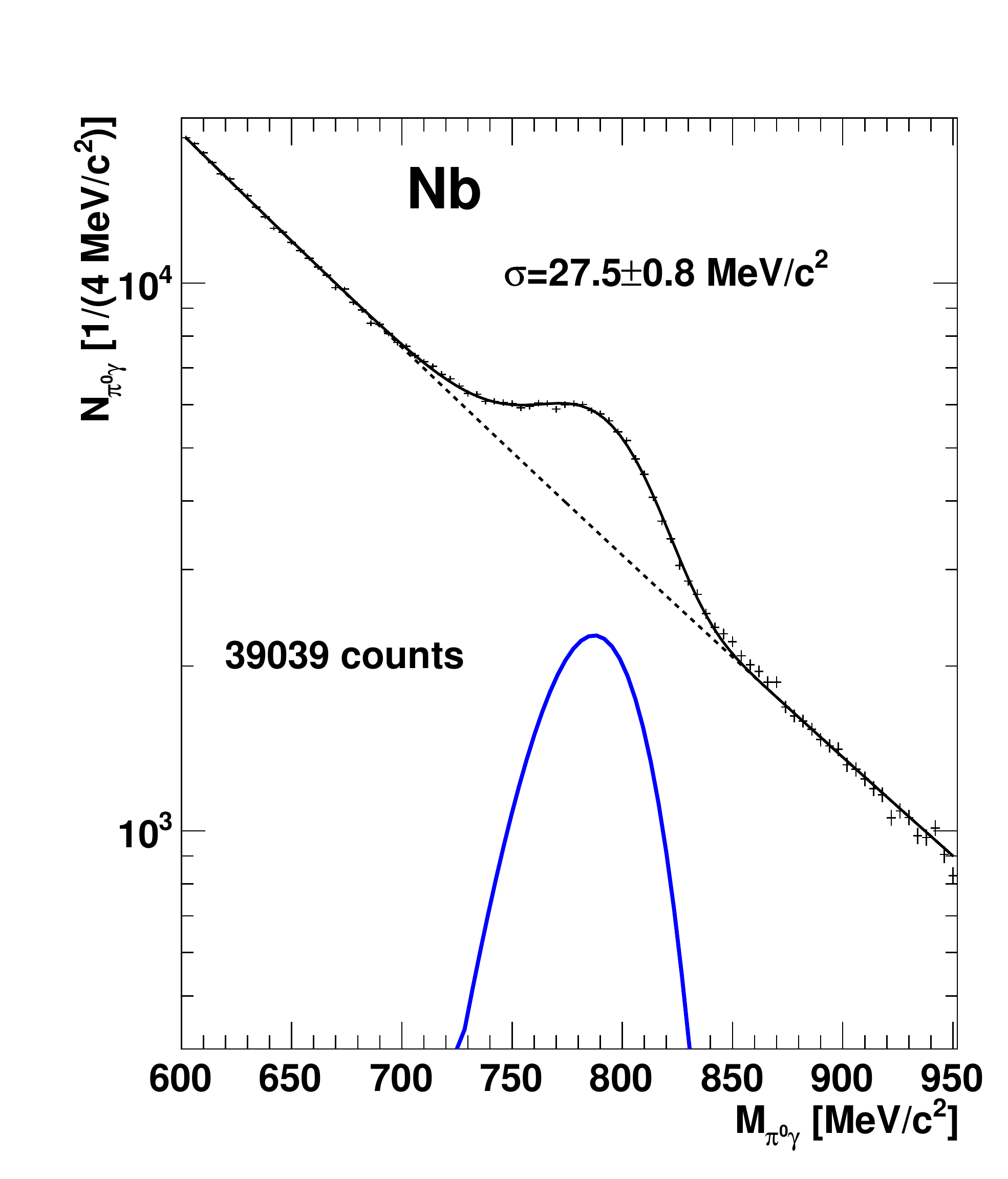} 
   \includegraphics{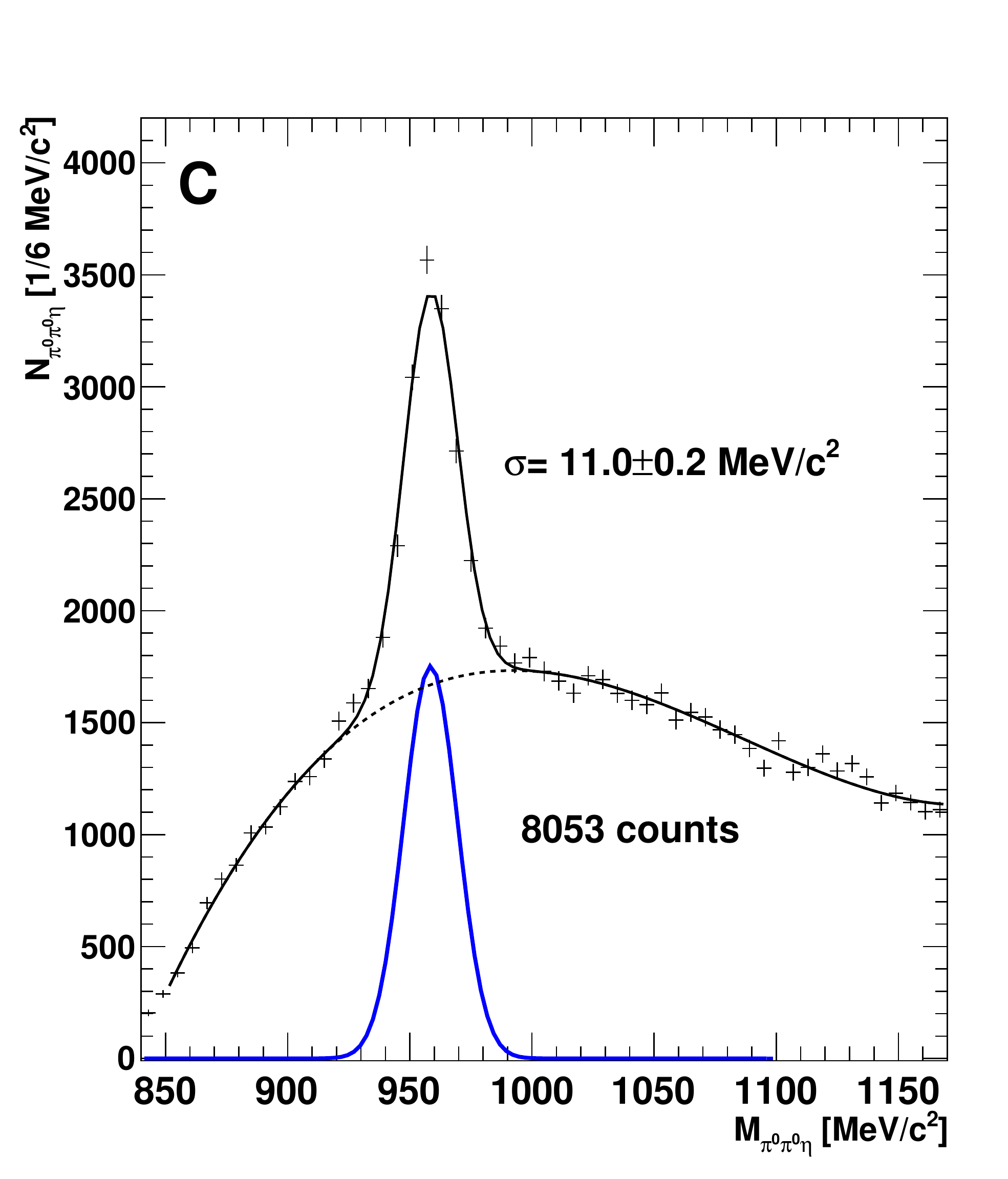}
   \includegraphics{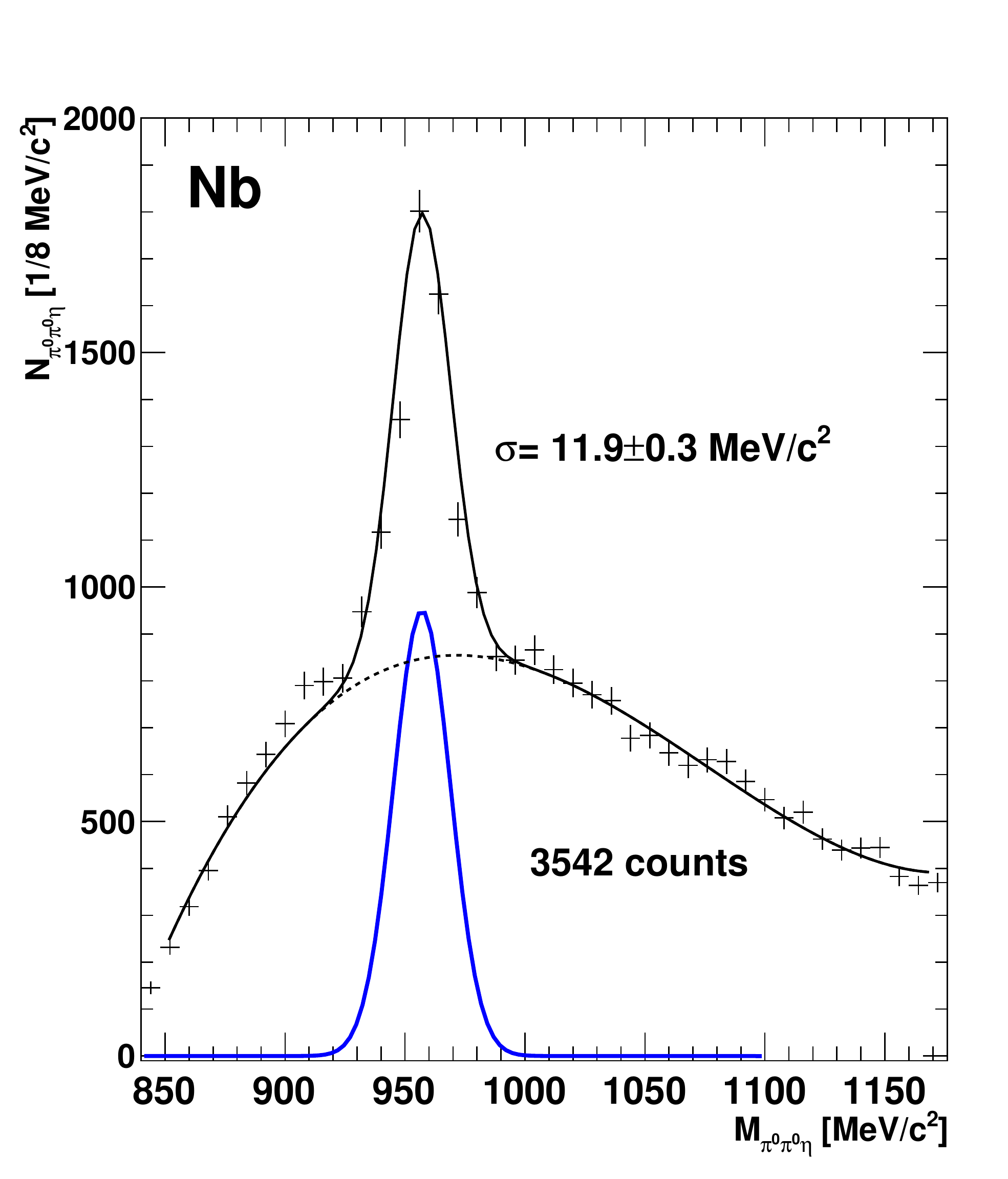}}
\caption{(Left) $\pi^0\gamma$ and (Right) $\pi^0\pi^0\eta$ invariant mass spectra obtained for the carbon and niobium target, respectively, and for incident photon energies of 1.2-2.9~GeV. The solid lines are fits to the invariant mass spectra using a Novosibirsk- \cite{MThiel} ($\omega$) and Gaussian- ($\eta^\prime$) line shape function together with an exponential ($\omega$) or polynomial ($\eta^\prime$), describing the background distribution. The $\sigma$ values correspond to the experimental mass resolution. The plots give the number of reconstructed mesons not corrected for photon flux, detection efficiency, and decay branching ratio.}
\label{fig:minv}
\end{center}
\end{figure*}
covering 96$\%$ of the full solid angle. In the angular range of 11$^\circ$-28$^\circ$ the CB modules (Forward Plug FP) were read out by photomultipliers, providing energy and time information while the rest of the CB crystals were read out by photodiodes with energy information only. Because of the high granularity and the large solid angle coverage the detector system was ideally suited for the detection and reconstruction of multi-photon events. Charged particles were registered in plastic scintillators in front of the MiniTAPS modules and the CB modules in the angular range of 11$^{\circ}$-28$^{\circ}$. In the polar angular range of 23$^{\circ}$-156$^{\circ}$ charged particles were identified in a three-layer scintillating fibre array.\\
To improve the statistics at low $\omega$ and $\eta^\prime$ momenta the diamond radiator was used in the niobium run to generate an excess of coherent photons peaking at an energy of 1.5~GeV in addition to the 1/E$_{\gamma}$  bremsstrahlung flux distribution. The polarisation of the radiation was not exploited in the analysis of the data. The photon flux through the target was determined by counting the photons reaching the gamma intensity monitor (GIM) at the end of the setup in coincidence with electrons registered in the tagging system. The total rate in the tagging system was $\approx$10~MHz. During the carbon run an aerogel-Cherenkov detector with a refractive index of n=1.05 was used to veto electrons, positrons and charged pions in the angular range covered by MiniTAPS. This device was replaced for the niobium beamtime with a gas-Cherenkov detector with a refractive index of n=1.00043 in order to veto electrons and positrons. The data were collected during two running periods of 525~h for the carbon and 960~h for the niobium target. 

The $\omega$ and $\eta^\prime$ mesons were identified via the $\omega \rightarrow \pi^{0} \gamma \rightarrow 3\gamma$ and $\eta^\prime\rightarrow \pi^0\pi^0\eta\rightarrow 6\gamma$ decay chains, which have a total branching ratio of 8.2$\%$ and 8.5$\%$, respectively \cite{PDG}. Events with $\omega $ and $\eta^\prime$ candidates were selected with suitable multiplicity trigger conditions. In the carbon run only events with at least four hits in the combined electromagnetic calorimeters were selected, requiring in addition that the aerogel-Cherenkov detector had not fired (veto-condition); in the niobium run a less restrictive trigger was applied, requiring two or more hits in the calorimeters and no hit in the gas-Cherenkov detector. The dead time introduced by the Cherenkov detectors were about 10$\%$ for the aerogel-Cherenkov detector and 25$\%$ for the gas-Cherenkov detector. The photon flux has been corrected for the GIM deadtime which was about 13$\%$ in the carbon run and 20$\%$ in the niobium run. A more detailed description of the detector setup and the running conditions can be found in \cite{Nanova_realC,AThiel}. 

 \section{Data analysis}
\label{sec:ana}
Events of interest were selected and the background suppressed by several kinematical cuts. Only events with incident photon energies larger than 1.2~GeV were processed. Photons were required to have energies larger than 25~MeV to suppress cluster split off. Random coincidences between the tagger and the detector modules in the first level trigger were removed by a cut in the corresponding time spectra around the prompt peaks and a sideband subtraction.
 \begin{figure*}
\begin{center}
 \resizebox{0.9\textwidth}{!}
 { \includegraphics{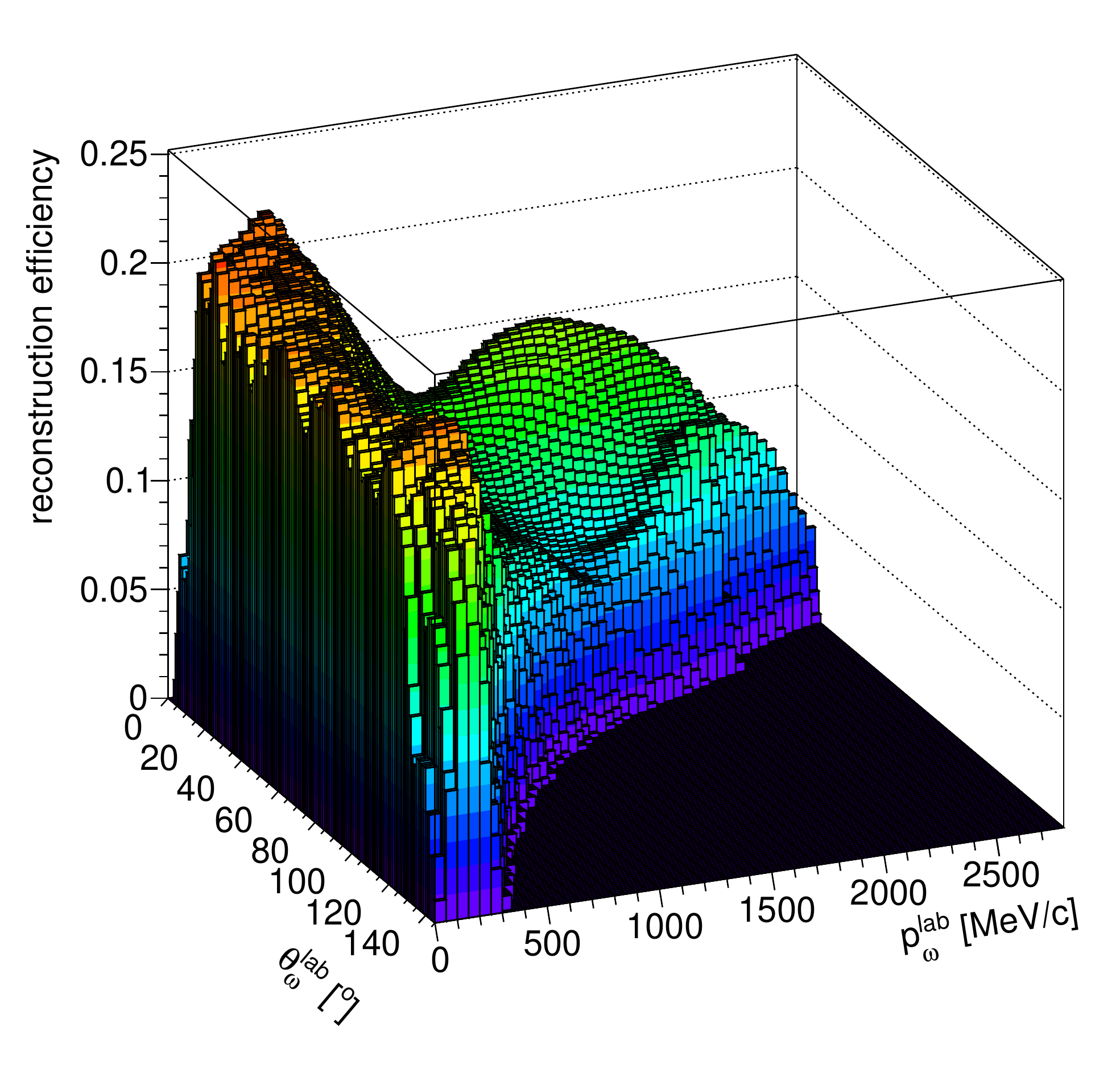}  \includegraphics{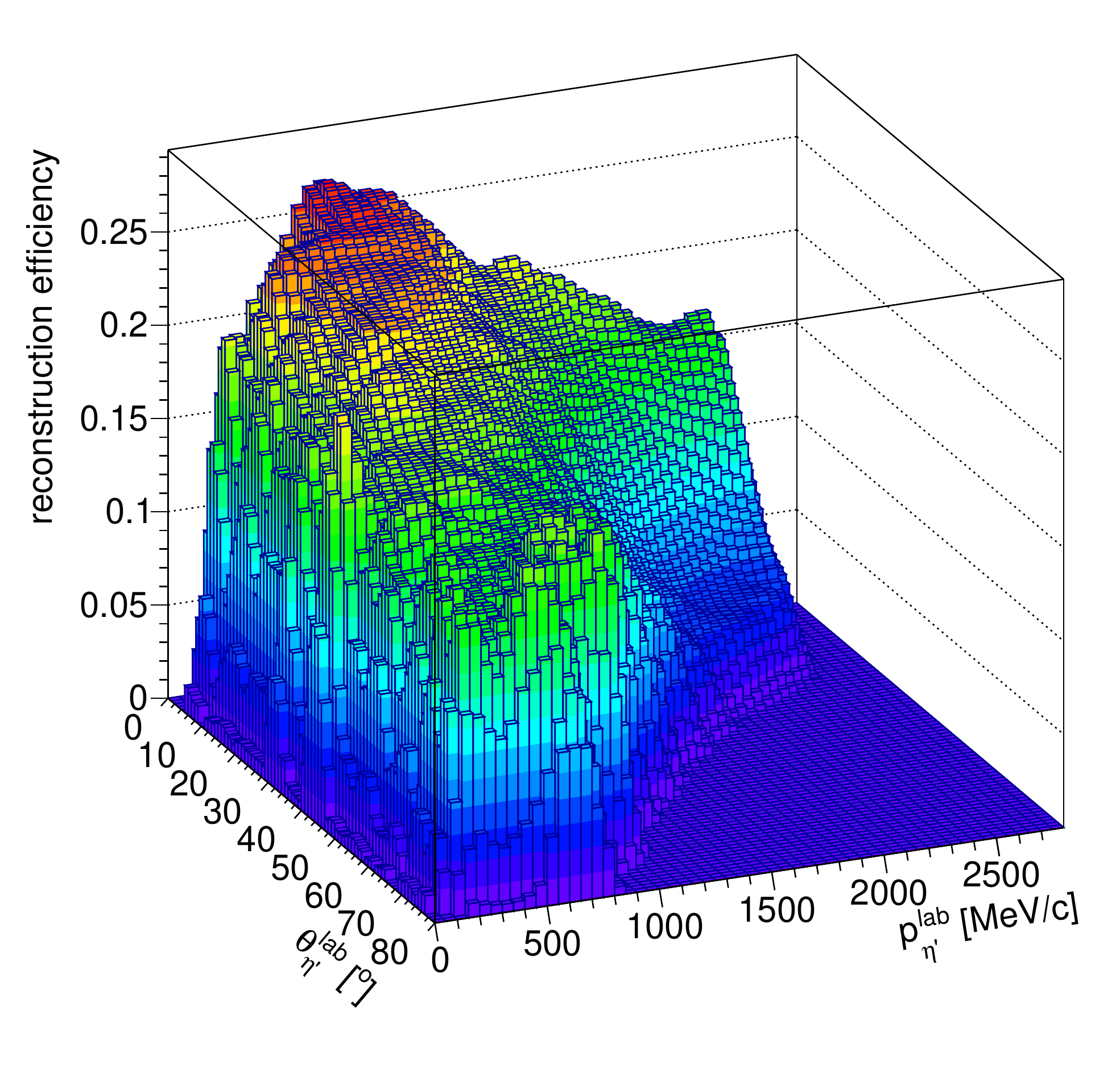}}
 \caption{Reconstruction efficiencies for (Left) the $p\omega $ final state with $\omega \rightarrow \pi^0 \gamma \rightarrow 3 \gamma$ and (Right) for $\eta^\prime$ mesons in the $\eta^\prime \rightarrow \pi^0 \pi^0 \eta \rightarrow 6 \gamma$ channel for photoproduction off carbon in the incident photon energy range of 1.2-2.9~GeV.} 
\label{fig:acc}
\end{center}
\end{figure*}

For the $\omega$ analysis, events with three photons and one charged hit were selected. The invariant mass of all photon pairs was calculated and the one combination closest to the $\pi^0$ mass of 135~MeV/$c^2$ was taken to be the $\pi^0$. Events with rescattered $\pi^0$ mesons from $\omega \rightarrow \pi^0 \gamma$ decays within the nucleus were suppressed by requesting the kinetic energy of the $\pi^0$ to be larger than 120~MeV \cite{Messchendorp,Kaskulov}.  Event losses due to this cut are taken into account in the simulation of the reconstruction efficiency (see below).The resulting $\pi^0\gamma$ invariant mass spectra for both targets are shown in Fig.~\ref{fig:minv} left.

For the $\eta^\prime$ analysis, events with only 6 photons and any number of charged hits and with an energy sum of neutral clusters larger than 600~MeV were selected. The 6 photons were combined in 2 pairs of 2 photons with invariant masses in the range 115~MeV/$c^2 \le m_{\gamma\gamma} \le$ 155~MeV/$c^2$ (corresponding to a $\pm$3$\sigma$ cut around $m_{\pi^{0}}$) and one pair with invariant mass in the range 510~MeV/$c^2 \le m_{\gamma\gamma} \le$ 590~MeV/$c^2$ (roughly corresponding to a $\pm$2$\sigma$ cut around $m_{\eta}$). The best photon combination was selected based on a $\chi^2$ minimization. To suppress the background from $\eta \rightarrow 3\pi^{0}$ decays and direct 3$\pi^{0}$ production, events with 3 $\gamma$ pairs with an invariant mass within the limits for the pion mass ($m_{\pi^{0}}$) given above were removed from the data set. The resulting $\pi^0\pi^0\eta$ invariant mass spectra for both targets are shown in 
Fig.~\ref{fig:minv} right.

For the determination of the momentum dependent differential cross sections the meson reconstruction efficiencies were determined by Monte Carlo simulations. In the event generator, the measured angular differential cross sections for $\omega$ \cite{Dietz} and  $\eta^\prime$ mesons \cite{Jaegle} produced off protons and neutrons bound in deuterium were used as input. The Fermi motion of nucleons in the target nucleus  was taken into account in the parametrisation proposed in \cite{Ciofi}. Photons from meson decays and the recoil nucleons emerging from the centre of the targets were tracked with the GEANT3 package \cite{GEANT} based on a full implementation of the detector system. 
A two-dimensional detector acceptance was determined as a function of the meson momentum and angle in the laboratory frame by taking the ratio of the number of reconstructed to the number of generated meson events for each momentum and angular bin. The resulting reconstruction efficiency distributions of $\omega$ and $\eta^\prime$ mesons produced off carbon are shown in Fig.~\ref{fig:acc}. 
The distributions vary smoothly as a function of the laboratory angle and momentum of the respective meson. In the Monte Carlo simulations the same trigger conditions as in the experiment were applied. Differences in the shape of the acceptance distributions for the two mesons are due to the fact that in case of the $\omega$ meson the multiplicity M=4 trigger condition can only be fulfilled if the three photons from the $\omega \rightarrow \pi^0 \gamma \rightarrow 3 \gamma$ decay and the recoil nucleon are detected while for the $\eta^\prime \rightarrow \pi^0 \pi^0 \eta \rightarrow 6 \gamma$ channel the recoil nucleon does not need to be registered to meet the trigger condition with M=4. Similar reconstruction efficiency distributions have been determined for the niobium target. The 2-dimensional reconstruction efficiency determination has the advantage that distortions of the meson angle and momentum due to final state interactions in the nucleus are directly taken into account by using the reconstruction efficiency for the observed final state meson momentum and angle.
When incrementing the $\pi^0 \gamma$ and  $\pi^0\pi^0\eta$ invariant mass histograms for the cross section determination, each event was weighted with the inverse photon flux at the given incident photon energy and the meson reconstruction efficiency for the observed meson momentum and angle in the laboratory frame (see Fig.~\ref{fig:acc}). The $\omega$ and $\eta^\prime$ yields were extracted by fitting the invariant mass spectra with a Novosibirsk- \cite{MThiel} ($\omega$) and Gaussian- ($\eta^\prime$) line shape function together with an exponential or polynomial to describe the background distribution. The invariant mass spectra were analysed for different 3-momenta of the $\pi^0 \gamma$ and $\pi^0 \pi^0 \eta$ pairs. The bin sizes were chosen according to the available statistics. The statistical errors, including the error of the random timing background subtraction, are accumulated in a separate histogram in parallel to the signal spectrum. The statistical errors are then obtained by summing over the same invariant mass range as used for fitting the signal.
\begin{figure*}
\begin{center}
 \resizebox{0.9\textwidth}{!}
 { \includegraphics{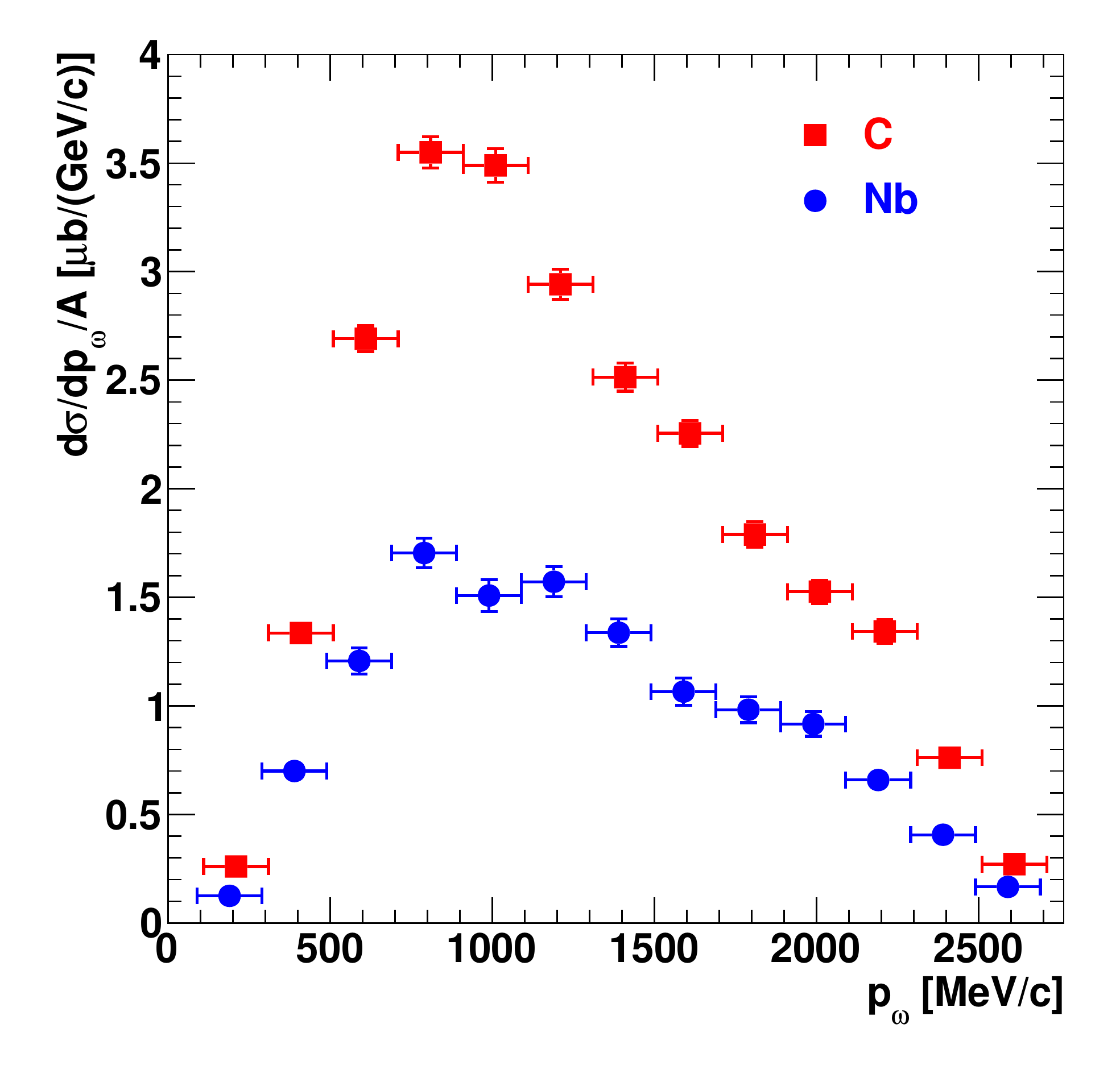}  \includegraphics{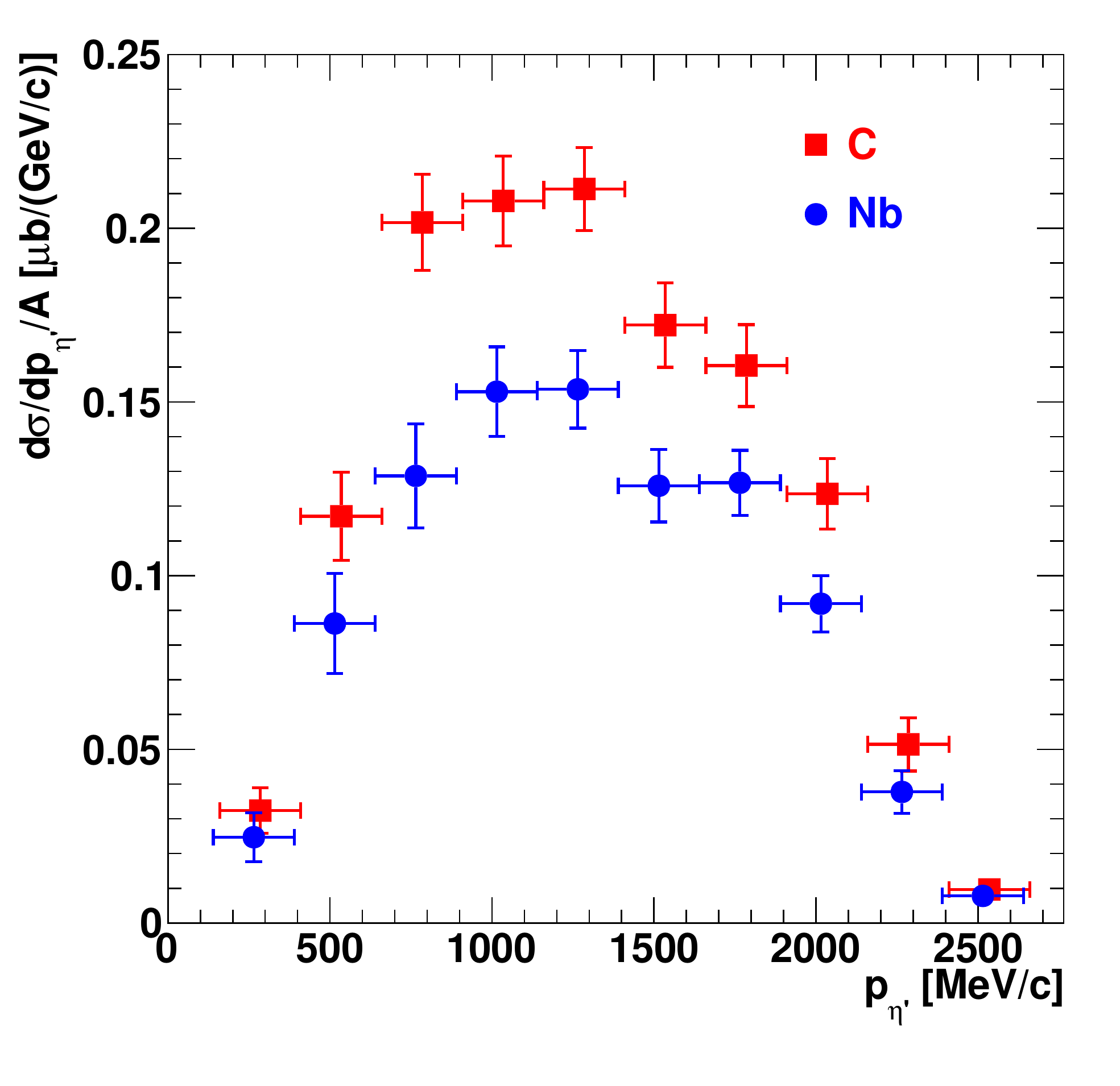}}
 \caption{Differential cross section per nucleon for (Left) $\omega$ and (Right) $\eta^\prime$ photoproduction off carbon (red squares) and niobium (blue circles) for the incident photon energies of 1.2-2.9~GeV. The data points for carbon are shifted by +10~MeV and for niobium by $-$10~MeV to avoid an overlap of the error bars.} 
\label{fig:dsigmadp}
\end{center}
\end{figure*}

The different sources of systematic errors for the cross section determination are summarized in Table \ref{tab:syst}. The systematic errors in the fit procedure were estimated to be in the order of 10$\%$ by applying different background functions and fit intervals. The systematic errors of the reconstruction efficiency were determined to be less than 10$\%$. This was estimated from changes in the acceptance in simulations assuming the extreme scenarios of isotropic and forward peaking $\omega$ and $\eta^\prime$ angular distributions.  Systematic errors associated with the photon flux determination using the GIM were estimated to be about 5-10$\%$. The systematic errors introduced by uncertainties in the photon shadowing \cite{Falter,bianchi,Muccifora} (see below) were $\approx 5\%$. Adding the systematic errors quadratically, the total systematic error of the cross section determinations was $\approx$ 17$\%$. 
\begin{table}[h!]
\centering
\caption{Sources of systematic errors for cross section determination.}
\begin{footnotesize}
\begin{tabular}{cc}
fits & $\approx$ 10$\%$\\
reconstruction efficiency & $\lesssim$10$\%$\\
photon flux & 5-10$\%$\\
photon shadowing & $\approx$ 5$\%$\\
\hline
total& $\approx$ 17$\%$\\
\end{tabular}
\end{footnotesize}
\label{tab:syst}
\end{table}
The determination of the transparency ratio discussed in the subsequent section requires the measurement of cross section ratios. Then, only the uncertainty for the reconstruction efficiency of the same meson in the same detector setup but for two different targets enters. This uncertainty is estimated to be reduced to 5$\%$. The systematic uncertainty for the cross section ratio is then 20$\%$.

\section{Experimental results}
Figure~\ref{fig:dsigmadp} presents the differential cross sections per nucleon for $\omega$ and $\eta^\prime$ photoproduction off carbon and niobium as a function of the meson momentum for incident photon energies of 1.2-2.9~GeV. Although the threshold for $\eta^\prime$ photoproduction off the free proton is E$_{\gamma}$=1.447~GeV an $\eta^\prime$ yield has been observed down to E$_{\gamma}$=1.2~GeV \cite{Nanova_realC} due to Fermi motion, broadening, and lowering of the $\eta^\prime$ mass. Thus the momentum distributions can be determined for $\omega$ and $\eta^\prime$ mesons over the same incident photon energy range. The cross sections include a 15$\%$ correction for absorption of the incoming photon beam (photon sha\-dow\-ing) for both nuclear targets \cite{Falter,bianchi,Muccifora}. The distributions show a maximum at around 800~MeV/$c$ and 1000~MeV/$c$ for the $\omega$ and $\eta^\prime$, respectively, and fall off towards higher momenta. The average momenta, given in Fig.~\ref{fig:dsigmadp}, are close to 1000~MeV/$c$.
\begin{figure*}
\begin{center}
 \resizebox{0.8\textwidth}{!}
 { \includegraphics{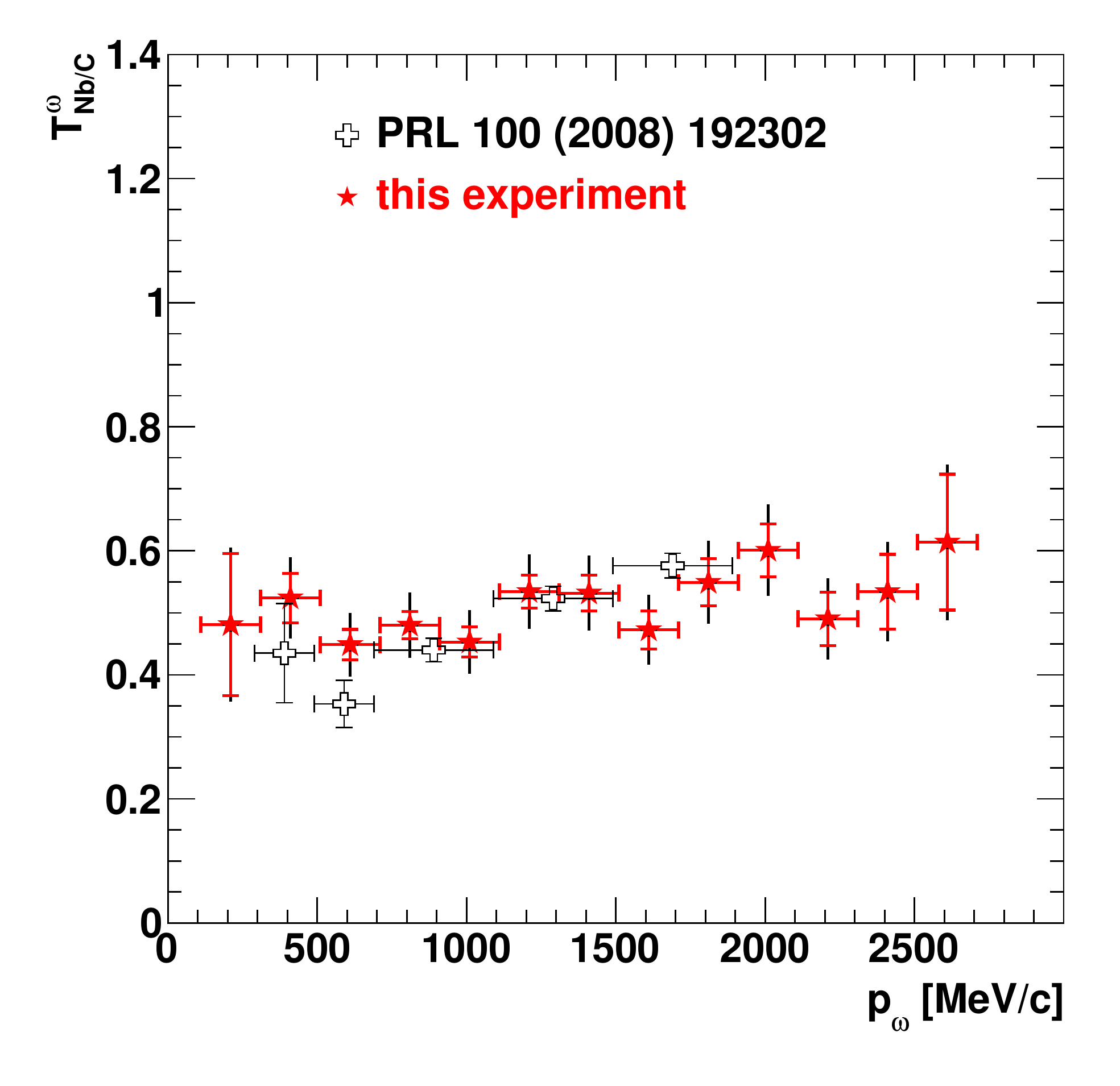}  \includegraphics{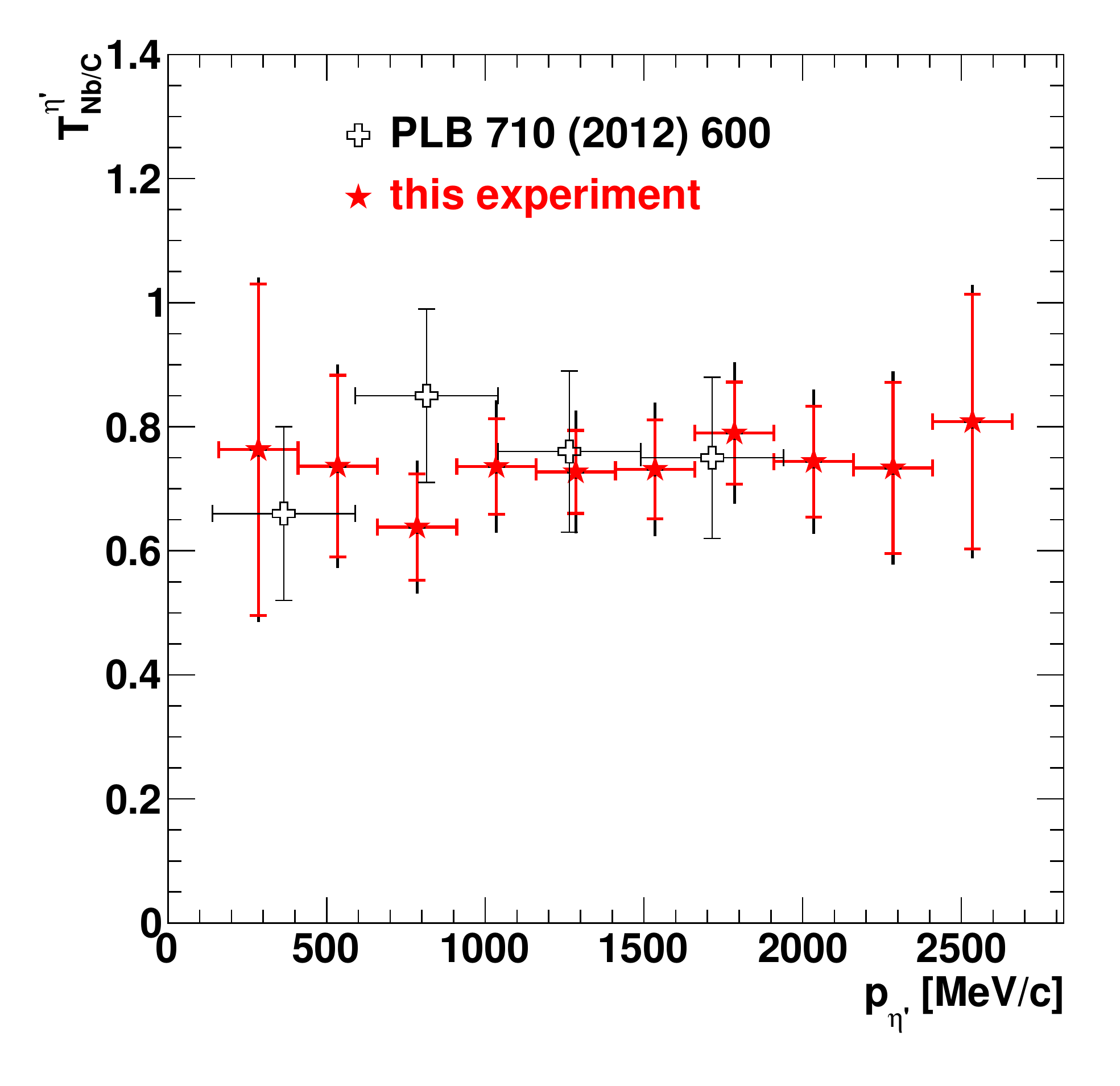}}
 \caption{The transparency ratio according to Eq. \ref{eq:trans} for (Left) $\omega$ and (Right) $\eta^\prime$ photoproduction off carbon and niobium (red stars) in comparison to earlier measurements (open crosses) \cite{Kotulla,Nanova_tr}. The data points of the present work are shifted by +10~MeV/$c$ and the previuosly published data by $-$10~MeV/$c$ to avoid an overlap of the error bars. For the present data the thick error bars (red) represent the statistical errors. The thin error bars (black) include the systematic errors added in quadrature.}
 \label{fig:TR}
\end{center}
\end{figure*}
To determine the attenuation of meson $m$ in nuclei and the inelastic meson-nucleon cross sections as a function of the meson momentum, the transparency ratio \cite{Cabrera} is deduced from the data:
\begin{equation}
T^\text{m}_{\text{A}}=\frac{ \sigma_{\gamma \text{A} \to \text{mX}}}{A \cdot \sigma_{\gamma \text{N} \to \text{mX}} }. \ \label{eq:trans1}
\end{equation} 
The meson production cross section per nucleon within a nucleus is compared to the production cross section off a free nucleon or off a proton or neutron bound in deuterium. Here, the nucleus serves as a target and at the same time as an absorber. If nuclei were completely transparent to the mesons the transparency ratio would be unity, as long as secondary production processes can be ignored. Although the photoproduction of $\omega$ and $\eta^\prime $ mesons off protons and neutrons bound in deuterium has been studied experimentally \cite{Jaegle,Dietz}, differential cross sections as a function of meson momentum are not available. The momentum dependence of the transparency ratio is thus obtained by dividing the differential inclusive meson production cross sections (see Fig.~\ref{fig:dsigmadp}) for niobium by the one for carbon. The transparency ratio is normalized to carbon according to
\begin{equation}
T^\text{m}_{\text{Nb/C}}=\frac{12\cdot \sigma_{\gamma \text{Nb} \to \text{mX}}}{93\cdot \sigma_{\gamma \text{C} \to \text{mX}} }, \ \label{eq:trans}
\end{equation} 
where 12 and 93 are the nuclear mass numbers of carbon and niobium, respectively. The normalization to a light nucleus like carbon has the further advantage that distortions due to two-body production- and absorption-processes are suppressed.

The resulting transparency ratios as a function of meson momentum are shown in Fig.~\ref{fig:TR}. Consistent with earlier measurements \cite{Kotulla,Nanova_tr}, a slight increase with momentum is observed for the $\omega$ meson while for the $\eta^\prime$ meson the transparency ratio is almost independent of momentum. Differences between the transparency ratios in the present measurements and the earlier ones reflect the systematic uncertainties of the measurements and are of the order estimated in section \ref{sec:ana}. For the present data the thick error bars (red) in Fig. \ref{fig:TR} and subsequent figures represent the statistical errors while the thinner error bars (black) include the systematic errors added in quadrature.

The interpretation of the transparency ratio in terms of meson absorption will only give reliable results if two-step production processes are negligible, where e.g. a pion is produced in an \begin{figure*}
\begin{center}
 \resizebox{0.8\textwidth}{!}
 { \includegraphics{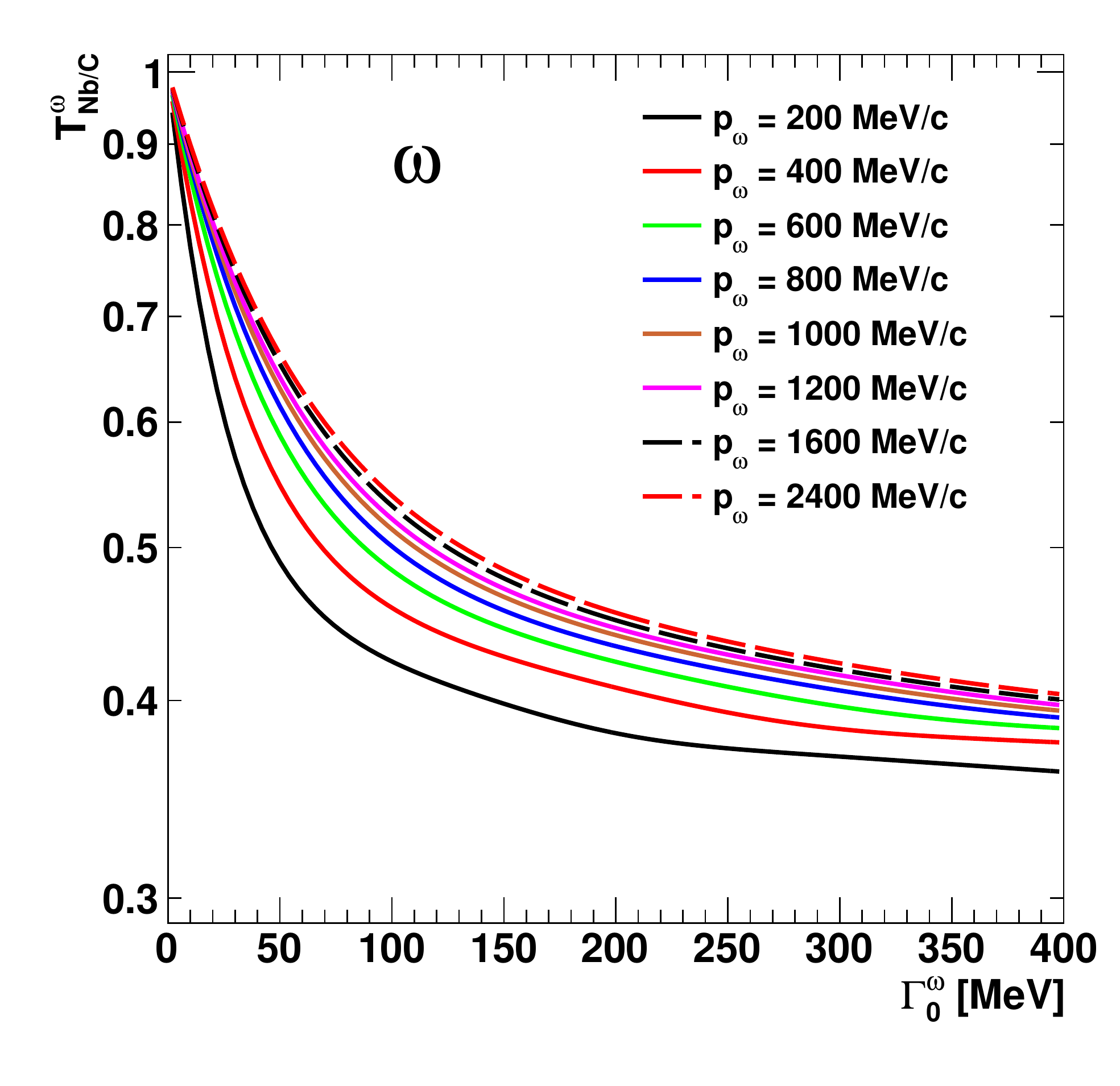}  \includegraphics{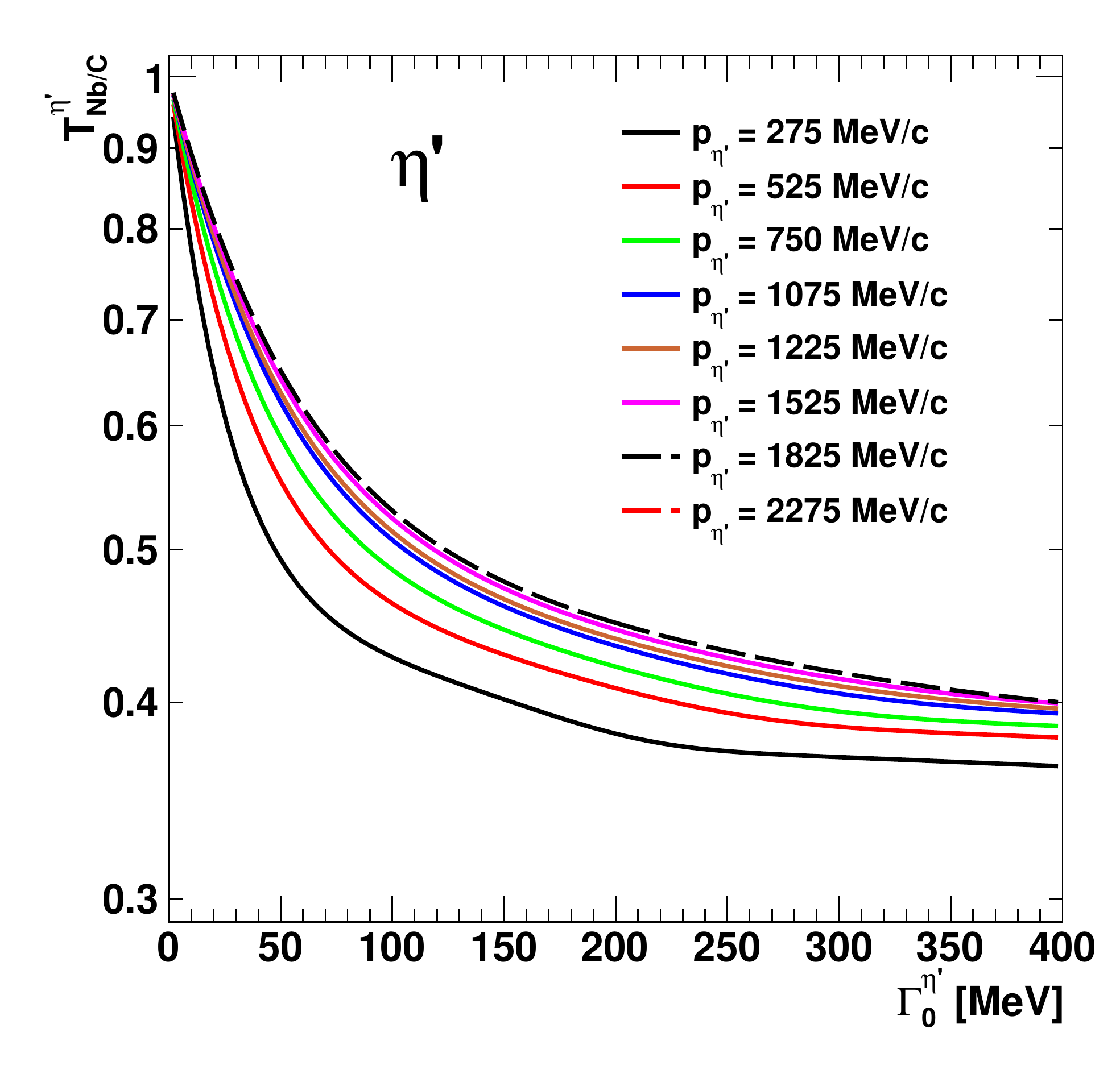}}
 \caption{The transparency ratio $T_{\text{Nb/C}}$ (Eq. \ref{eq:trans}) calculated as described in the text for different values of the width $\Gamma_0$ as a function of the (Left) $\omega$ and (Right) $\eta^\prime$ momentum in the nuclear restframe.}
 \label{fig:TR_vs_Gamma}
\end{center}
\end{figure*}
initial step followed by production of the meson of interest in a subsequent pion-induced reaction on another nucleon within the nucleus. As shown in \cite{Nanova_tr}, two-step processes are negligible for $\omega$ and $\eta^\prime $ mesons since the spectral distribution of pions falls off towards higher energies and - at the required pion momenta of $\approx$ 1.3 and 1.5~GeV/$c$ - pion induced meson production cross sections are only $\approx$ 2.5~mb and 0.1~mb, respectively, compared to the total reaction cross section of about 30-40 mb \cite{PDG}. Consequently, two-step processes are neglected in the subsequent analysis of the transparency ratio.

Following \cite{Kotulla,Paryev,Nanova_tr} the in-medium meson width is deduced from the measured transparency ratio within a Glauber model in the high energy eikonal approximation according to 
\begin{equation}
T_\text{A} = \frac{2\pi}{A} \int_{r_{\perp}=0}^R \int_{z=-\sqrt{R^2-r_{\perp}^2}}^{\sqrt{R^2-r_{\perp}^2}} \! \mathrm{d}r_{\perp} \mathrm{d}z \space\  r_{\perp} \cdot \rho(z,r_{\perp}) \cdot att(z,r_{\perp}) .  \label{eq:T_A}
\end{equation}
Here, $att(z,r_{\perp})$ is the attenuation of a meson produced at position $(z,r_{\perp})$ within the nucleus:
\begin{equation}
 att(z,r_{\perp})= exp \Big\{-\frac{E\cdot\Gamma_0}{\hbar c \cdot \rho_0} \int_{x=z}^{\sqrt{R^2-r_{\perp}^2}} \! \mathrm{d}x \frac{\rho(x,r_{\perp})}{p^*(x,r_{\perp}) c}\Big\} \label{eq:att}
\end{equation}
For the nuclear density profile $\rho(r)$ of carbon and niobium a harmonic oscillator distribution \cite{Nieves} and a two-parameter Fermi distribution \cite{Jager} have been chosen, respectively.
The distance R from the centre of the nucleus, where the nuclear density has dropped to $<$ 0.2$\%$ of the nuclear saturation density $\rho_0$, is 5~fm for carbon and 8~fm for niobium. The in-medium width $\Gamma_0$ of the meson at density $\rho_0$ refers to the nuclear restframe. Hereby, it is assumed that the produced mesons go dominantly in forward direction on straight line trajectories. This assumption appears justified in view of the observed strong forward rise in the $\omega$ angular distribution (t-channel production)
  \cite{Dietz}. For the $\eta^\prime$ meson this forward peaking is not as pronounced but still significant \cite{Nanova_realC}. 

In the above derivation the width $\Gamma(\rho) $ is assumed to depend linearly on the nuclear density $\rho$:
\begin{equation}
\Gamma(\rho) = \Gamma_0 \cdot \frac{\rho}{\rho_0}.\label{eq:Gamma},
\end{equation} 
where $\Gamma_0$ is the momentum dependent width at normal nuclear matter density. The in-medium meson momentum $p^*(x,r_{\perp})$ is given by
\begin{equation}
p^*(x,r_{\perp}) c= \sqrt{E^2-(m^*(x,r_{\perp})c^2)^{2}}\label{eq:p^*}
\end{equation}
with
\begin{equation}
m^*(x,r_{\perp})c^2 = m_0 c^2+ V_0 \cdot \frac{\rho(x,r_{\perp})}{\rho_0}\label{eq:m^*}.
\end{equation}
Hereby, E is the total energy of the meson
\begin{equation}
E=\sqrt{(m_0c^2)^2+(p_0c)^2} = \sqrt{(m^*c^2)^2 + (p^*c)^2} \label{eq:E}
\end{equation}
and $m^*,m_0,p^*,p_0$ are the meson mass and momentum inside and outside of the nuclear medium, respectively. 

Using Eqs. \ref{eq:T_A}-\ref{eq:E} the escape probability of $\omega$ and $\eta^\prime$ mesons from carbon and niobium has been calculated for different meson momenta and in-medium widths $\Gamma_0$. Taking the ratio of the meson escape probabilities, the transparency ratio $T_{\text{Nb/C}}$ is deduced and plotted in Fig.~\ref{fig:TR_vs_Gamma} as a function of $\Gamma_0$ for different meson momenta.
From the measured transparency ratio $T_{\text{Nb/C}}$ and any given meson momentum the in-medium width $\Gamma_0$ is extracted from the calculated curves in 
Fig.~\ref{fig:TR_vs_Gamma}. In this analysis the modification of the in-medium meson mass $m^*$ and momentum $p^*$  by the real part of the meson-nucleus potential is taken into account. In the calculations, potential depths $V_0$ at normal nuclear matter density of $-$30 and $-$40~MeV are used for the $\omega$ and $\eta^\prime$ meson, respectively, as recently determined in \cite{Nanova_realC,Metag_Hypint,Friedrich}. The effect on the calculated transparency is, however, very small. For meson momenta $>$~400~MeV/$c$ the change in the transparency ratio is less than 2$\%$ compared to the case $V_0$ = 0 and less than 10$\%$ for the smallest widths and meson momenta.

\begin{figure*}
\begin{center}
 \resizebox{1.0\textwidth}{!}
 { \includegraphics{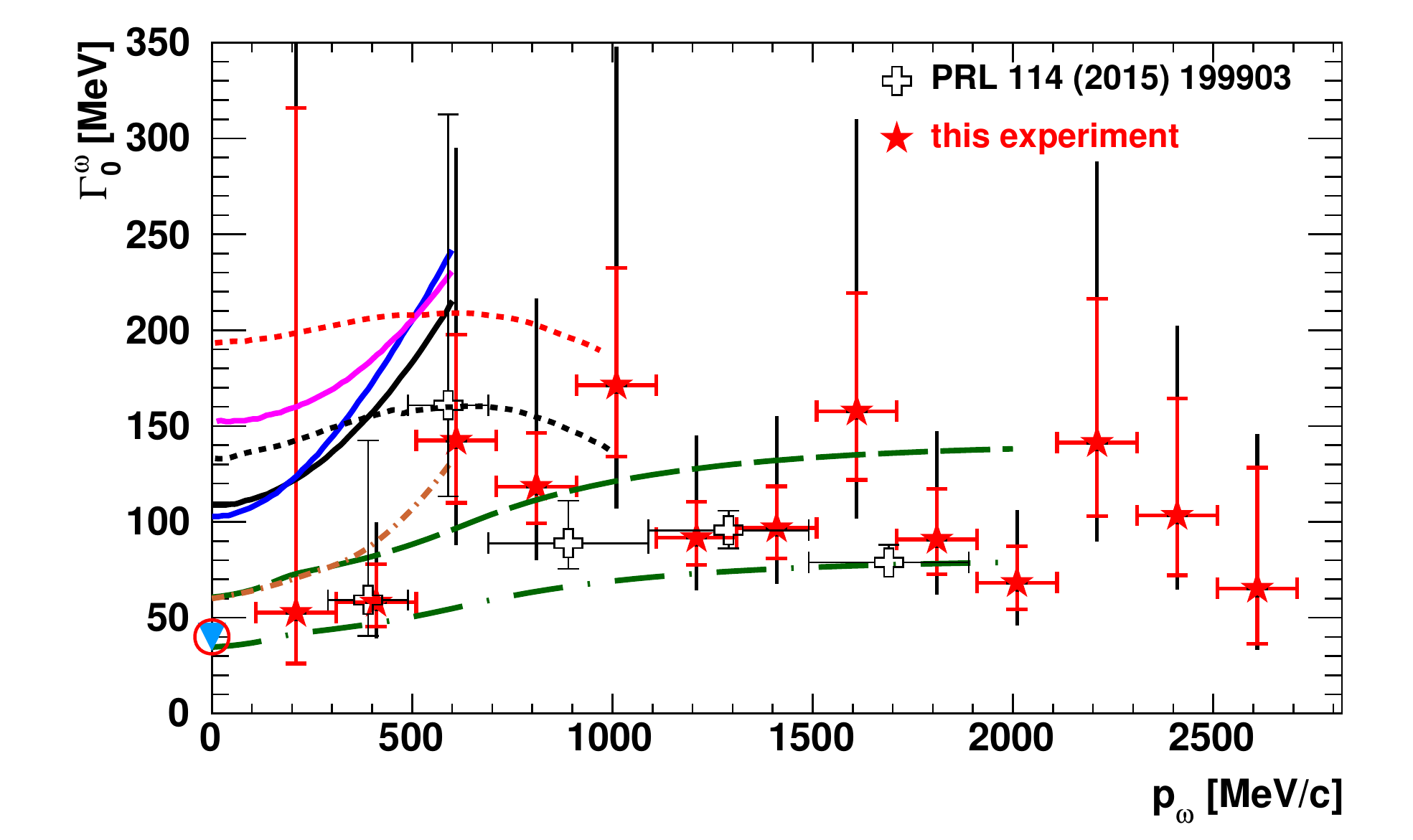}  \includegraphics{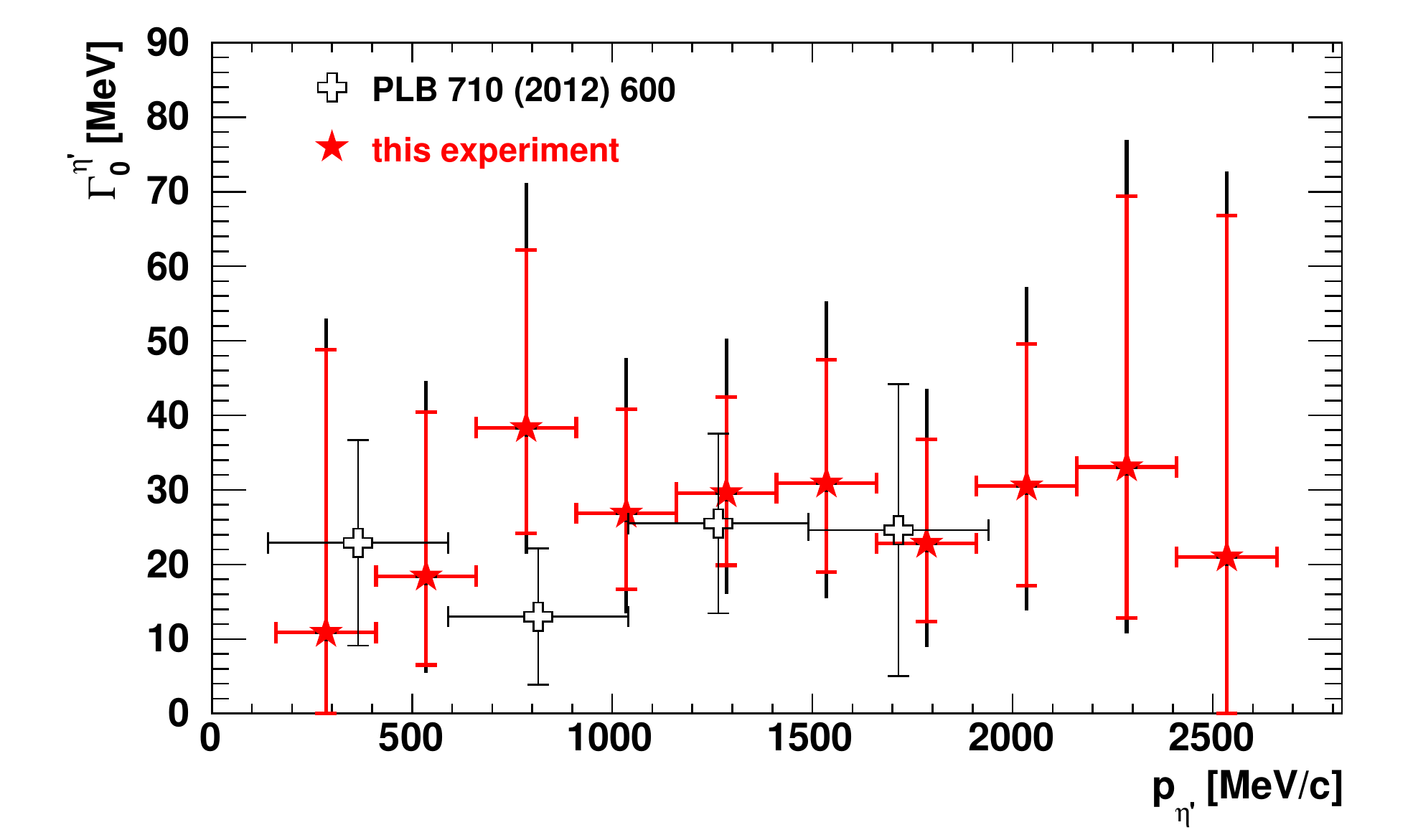}}
 \caption{In-medium width $\Gamma_0$ of (Left) $\omega$ and (Right) $\eta^\prime$ mesons as a function of the meson momentum (red stars), derived from the data presented in Fig.~\ref{fig:TR} using the curves of Fig.~\ref{fig:TR_vs_Gamma}, in comparison to earlier measurements (open crosses) \cite{Kotulla,Kotulla_err,Nanova_tr}. The data points of the present work are shifted by +10~MeV/$c$ and the previously published data by $-$10~MeV/$c$ to avoid an overlap of the error bars (symbols as in Fig. \ref{fig:TR}). The solid curves correspond to calculations by Ramos \textit{et al.} \cite{Ramos}, the short dashed curves to calculations by Ca$\-$brera and Rapp \cite{Cabrera_Rapp} for different model assumptions (see text). The brown dashed-dotted curve shows the momentum dependence of the $\omega$ in-medium width calculated in a coupled-channel resonance model \cite{Muehlich}. The long dashed green curves correspond to different options for the in-medium $\omega$ width used in GiBUU simulations \cite{Buss}. The blue triangle and the open red circle represent the widths calculated for an $\omega$ meson at rest in the nuclear medium in \cite{Klingl,Lutz}, respectively.}
 \label{fig:gamma}
\end{center}
\end{figure*}

As a result, the in-medium width $\Gamma_0$ is shown in Fig.~\ref{fig:gamma} as a function of the $\omega$ and $\eta^\prime$ momentum, respectively. The non-linearity in the  correlation between transparency ratio and in-medium width (see  Fig.~\ref{fig:TR_vs_Gamma}) introduces strongly asymmetric error bars. Within errors, the results are consistent with previous measurements \cite{Kotulla,Kotulla_err,Nanova_tr}.

It is immediately apparent that the $\omega$ widths are larger than the $\eta^\prime$ widths by about a factor three. The finer binning in the present data also reveals more clearly a variation of the widths with momentum, indicating a rise with momentum at small momenta and a fall-off towards higher momenta. 

The data for the $\omega$ meson are compared to calculations of the in-medium $\omega$ width. Cabrera and Rapp \cite{Cabrera_Rapp} and Ramos \textit{et al.} \cite{Ramos} have studied the width of the $\omega$ meson in cold nuclear matter as a function of the nuclear density and the meson 3-momentum. Both groups independently find that the main contribution to the in-medium $\omega$ width is determined by the $\omega \to \rho \pi$ channel whereby the dressing of the $\pi$ and $\rho$ propagator in the medium is essential. Thus, the $\pi\rho$ cloud is the main agent for the in-medium broadening of the $\omega$ meson. Both groups obtain in-medium $\omega$ widths of the order of 100-200~MeV, somewhat larger than observed experimentally. They differ in the partitioning into the $\pi$ and $\rho$ modifications and obtain differences in the 3-momentum dependence of the $\omega$ width. While Cabrera and Rapp \cite{Cabrera_Rapp} find a moderate momentum dependence, Ramos \textit{et al.} \cite{Ramos} get an almost linear increase of the width with momentum. Alternatively, Klingl et al. \cite{Klingl}, Lutz et al. \cite{Lutz} and M\"uhlich et al. \cite{Muehlich} consider the coupling of the $\omega$ meson to nucleon resonances as the main effect driving the in-medium broadening of the $\omega$ meson. In fact, the data are closer to the momentum dependence of the $\omega$ in-medium width calculated in a coupled-channel resonance model \cite{Muehlich} and are only slightly larger at low momenta than the width of 40 MeV calculated for the $\omega$ at rest in the nuclear medium \cite{Klingl,Lutz}. The statistics of the present experiment is unfortunately not sufficient to clearly discriminate between these different theoretical approaches. 
\begin{figure*}
\begin{center}
 \resizebox{1.0\textwidth}{!}
 { \includegraphics{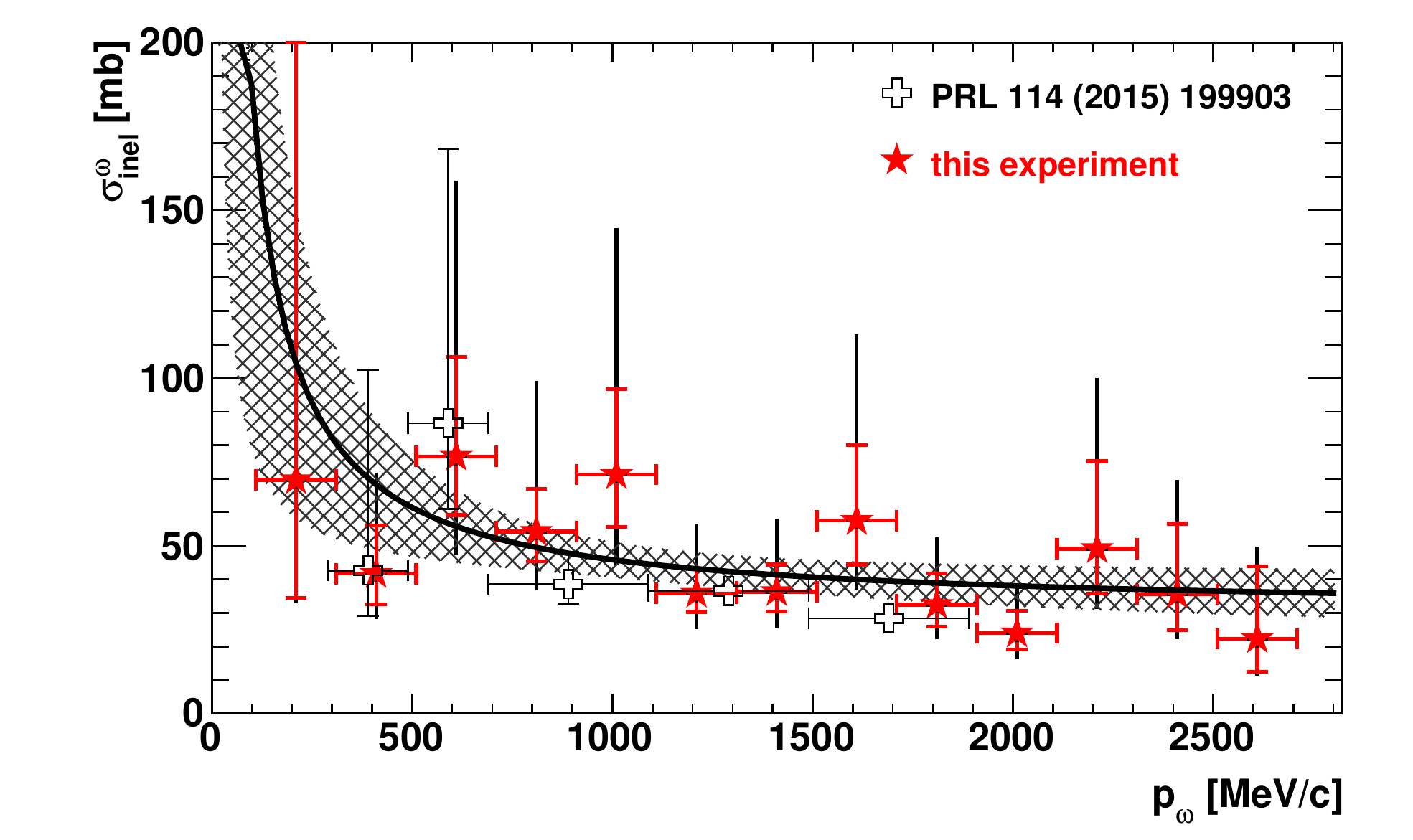}  \includegraphics{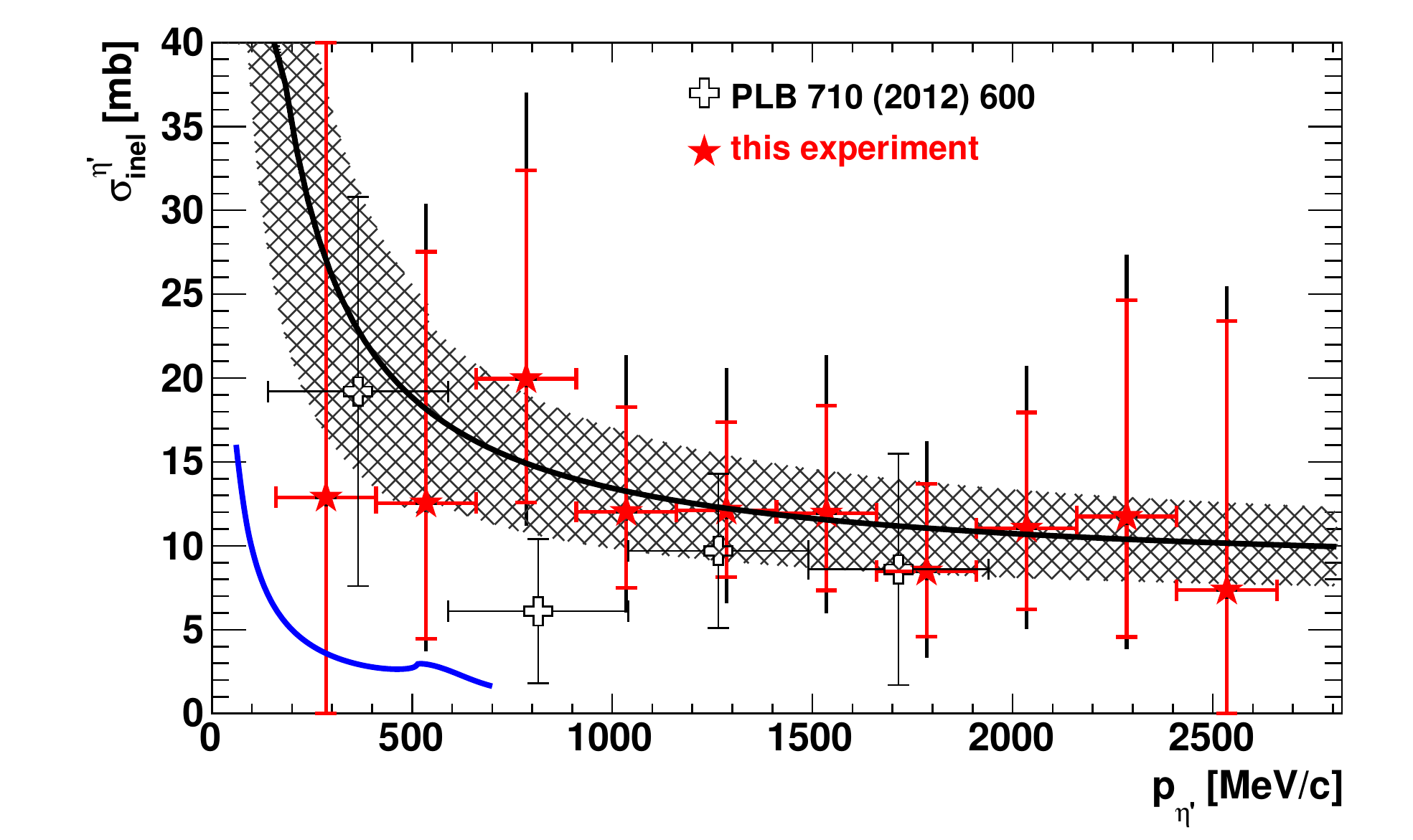}}
 \caption{Inelastic $\omega$-nucleon (Left) and $\eta^\prime$-nucleon (Right) cross sections deduced from Eq. \ref{eq:Gamma-sigma} as a function of the meson momentum (red stars) in comparison to earlier measurements (open crosses) \cite{Kotulla,Kotulla_err,Nanova_tr}. The data points of the present work are shifted by +10~MeV/$c$ and the previously published data by $-$10~MeV/$c$ to avoid an overlap of the error bars (symbols as in Fig. \ref{fig:TR}). The solid black curve is a fit to the data of this experiment using the parametrisation $\sigma_{\text{inel}}[mb] = a + \frac{b}{p[\mathrm{GeV}/c]}$\cite{Lykasov} with $a$ = 15.5 $\pm$ 15.1 and $b$ = 30.3 $\pm$ 13.8 for the $\omega$ meson and $a$ = 5.4 $\pm$ 8.1 and $b$ = 8.0 $\pm$ 7.0 for the $\eta^\prime$ meson (the parameters a,b are strongly anticorrelated). The shaded areas indicate a confidence level of $\pm$1$\sigma$ of the fit curve taking statistical and systematic errors into account. The error weighted mean value of the $\eta^\prime$ absorption cross section is (13 $\pm$ 3) mb. The blue curve represents the inelastic $\eta^\prime$-nucleon cross section calculated in \cite{Oset_Ramos}.  }
 \label{fig:sig_inel}
\end{center}
\end{figure*}

\begin{figure*}
\begin{center}
 \resizebox{1.0\textwidth}{!}
 { \includegraphics{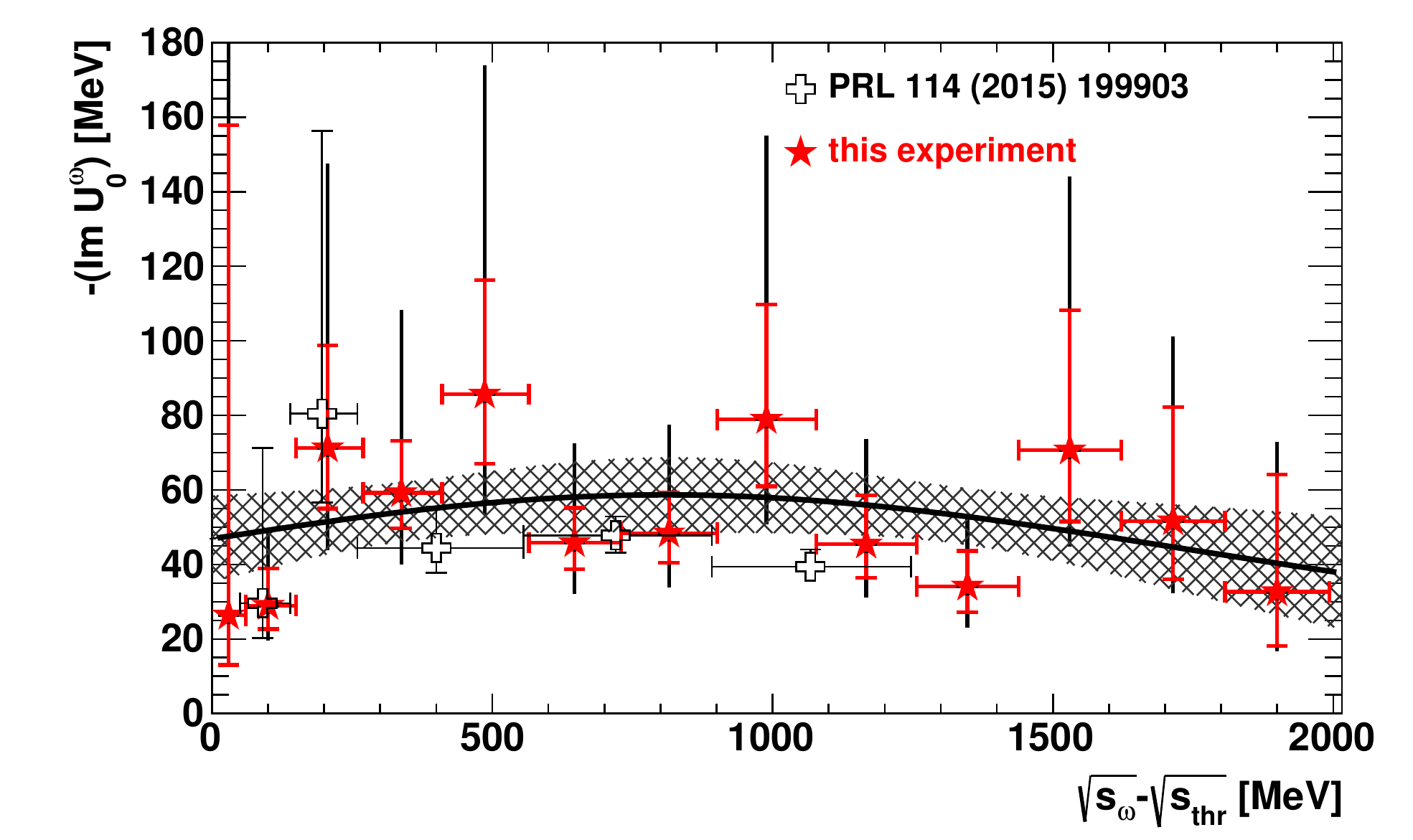}  \includegraphics{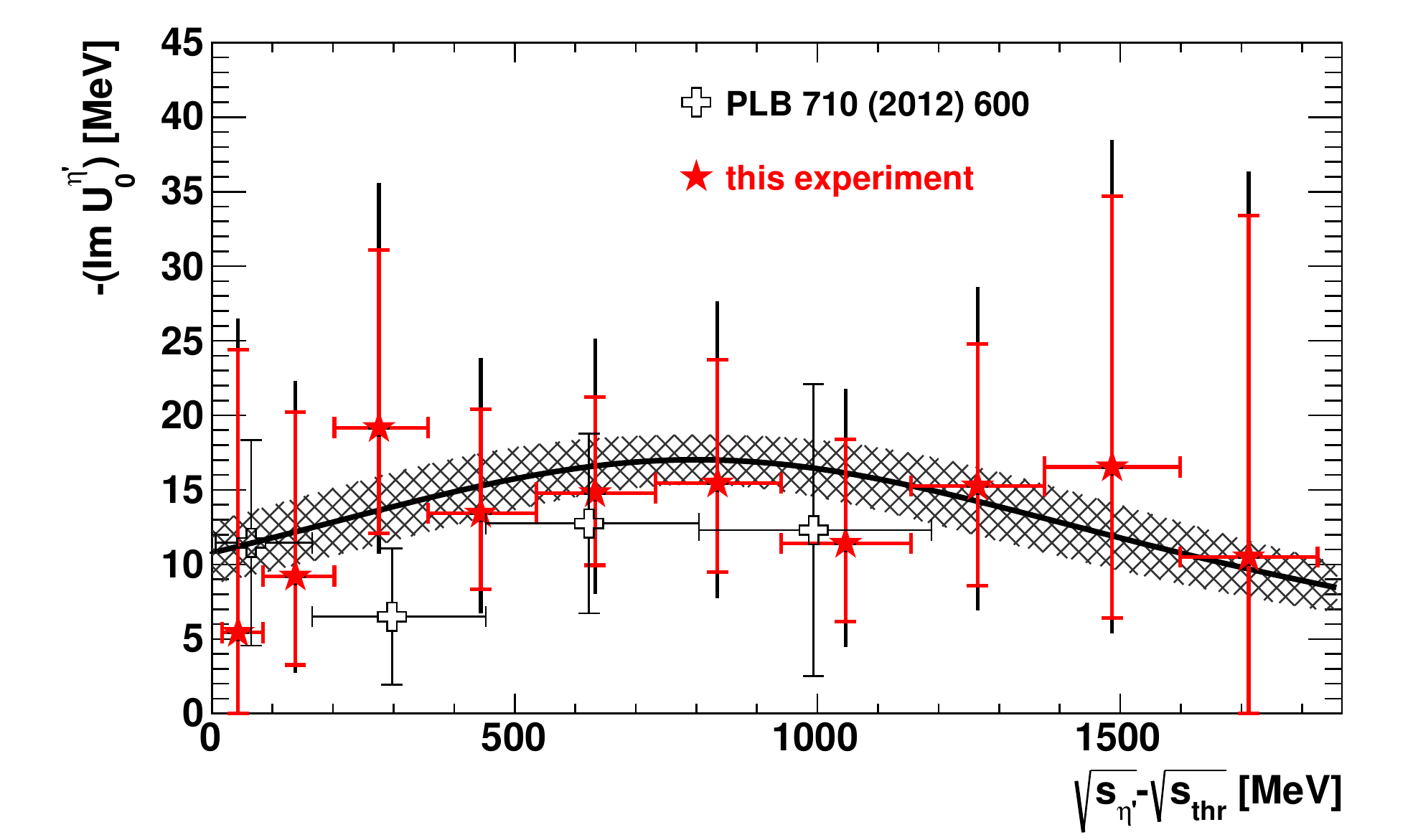}}
 \caption{Imaginary part of the (Left) $\omega$-nucleus and (Right) $\eta^\prime$-nucleus optical potential as a function of the available energy in the meson-$^{93}$Nb system (red stars) in comparison to earlier measurements (open crosses) \cite{Kotulla,Kotulla_err,Nanova_tr}. The data points of the present work are shifted by +5~MeV and the previuosly published data by $-$5~MeV to avoid an overlap of the error bars (symbols as in Fig. \ref{fig:TR}). The solid curves are Breit-Wigner fits to the present data. The shaded areas indicate a confidence level of $\pm$1$\sigma$ of the fit curve taking statistical and systematic errors into account.} 
 \label{fig:ImU}
\end{center}
\end{figure*}

From the in-medium widths $\Gamma_0$ of Fig.~\ref{fig:gamma} inelastic cross sections $\sigma_{\text{inel}}$ can be derived, using Eq. \ref{eq:Gamma-sigma}. 
In the low-density approximation the in-medium meson width $\Gamma (\rho=\rho_0) = \Gamma_0$ at normal nuclear matter density $\rho_0$ and the inelastic meson-nucleon cross section $\sigma_{\text{inel}}$ are related by
\begin{equation}
\Gamma (\rho=\rho_0) = \Gamma_0 = \hbar c \cdot \rho_{0} \cdot \sigma_{\rm inel} \cdot p*/E \label{eq:Gamma-sigma}.
\end{equation}
\par
The resulting inelastic cross sections are shown in Fig.~\ref{fig:sig_inel} as a function of the meson momentum. To compare the data to a parametrization frequently used in the literature, the data for both mesons have been fitted with an ansatz 
\begin{equation}
\sigma_{\text{inel}}[mb]= a +  \frac{b}{p[\mathrm{GeV}/c]} \label{eq:sigma},
\end{equation}
as proposed by Lykasov \textit{et al.} \cite{Lykasov} and used as parametrisation in GiBUU transport simulations \cite{Buss}. The present inelastic $\eta^\prime$ cross section data (Fig.~\ref{fig:sig_inel} (right)) shows a mean value of (13~$\pm$~3) mb, slightly larger but consistent with the earlier result of (10.3 $\pm$ 1.4) mb reported in \cite{Nanova_tr}. The experimental data are compared to calculations by Oset and Ramos \cite{Oset_Ramos}. They have studied the $\eta^\prime$-nucleon interaction within a chiral unitary approach, including $\pi N$ and $\eta N$ coupled channels, which yields a very weak $\eta^\prime N$ interaction. The $\eta^\prime N$ amplitude is substantially enhanced when vector meson-baryon states are included in the coupled channel scheme via normal and anomalous couplings of pseudo-scalar to vector mesons. In this approach inelastic $\eta^\prime N$ cross sections rising from about 3 mb at $p_{\eta^\prime} $ = 600~MeV/$c$ to about 20 mb at $p_{\eta^\prime}$ = 50~MeV/$c$ are predicted. The calculations seem to underestimate the experimentally determined inelastic $\eta^\prime$ cross section. This may not be surprising since multi-particle production, probably dominant because of the large $\eta^\prime$ mass, has not been considered in \cite{Oset_Ramos}.

\par
As a final step, the momentum dependence of the in-medium $\omega$ and $\eta^\prime$ widths from Fig.~\ref{fig:gamma} can be converted into the dependence of the imaginary part of the $\omega$- and $\eta^\prime$-nucleus potential as a function of the available energy in the meson-$^{93}$Nb system, as shown in Fig.~\ref{fig:ImU}. The imaginary part of the potential $\text{Im} U$ at normal nuclear matter density is just half of the in-medium width $\Gamma_0$ (see Fig.~\ref{fig:gamma}). The finer binning of the present data allows a more reliable extrapolation towards the production threshold by fitting the data. Several fit functions have been applied (polynomial of 1st. and 2nd. order, Gaussian, Breit-Wigner). The range of -$\text{Im} U(0)$ values obtained for different fit functions reflects the systematic uncertainties. For the $\omega$ meson the modulus of the imaginary part of the meson nucleus potential near threshold is found to be (48~$\pm$~12(stat)$\pm$~9(syst))~MeV comparable to the modulus of the real part of about 30~MeV, determined in \cite{Metag_PPNP,Metag_Hypint,Friedrich}. For the $\eta^\prime$ meson the extrapolation towards the production threshold yields an imaginary potential of (13 $\pm$ 3(stat)$\pm$3(syst)) MeV, corresponding to an imaginary part of the $\eta^\prime$ scattering length $\text{Im}(a_{\eta^\prime N}$) = (0.16 $\pm$ 0.05)~fm. This is about a factor two smaller than obtained in the direct determination of the $\eta^\prime N$ scattering length from an analysis of near threshold $\eta^\prime$ production in the $p p \rightarrow p p \eta^\prime$ reaction \cite{Moskal}. The error bars of both completely independent determinations do, however, almost overlap. The imaginary part of the meson nucleus potential is about 3 times smaller than the real part of about 40~MeV \cite{Nanova_realC}. As a consequence, the $\eta^\prime$ meson is a good candidate for the search for meson-nucleus bound states. Although the real potential is not very deep, the small imaginary potential may allow for the existence of relatively narrow bound states. On the other hand, the $\omega$ meson is not a good candidate for the search for meson-nucleus bound states as the width of these states is expected to be very large which makes it difficult to detect them experimentally.

As discussed in the introduction, the energy dependence of the imaginary part of the $\omega$ and $\eta^\prime$ nucleus potential (Fig.~\ref{fig:ImU}) may serve as input for a dispersion relation analysis. Such an analysis would allow for a consistency check of the real and imaginary part of the meson nucleus potential determined independently in different experiments~\cite{Lenske}.

\section{Conclusions}
Differential cross sections as a function of the meson momentum have been measured in photoproduction of $\omega$ and $\eta^\prime$ mesons off carbon and niobium. Based on these cross sections the momentum dependence of the Nb/C transparency ratio has been deduced. Within a Glauber model analysis in the high energy eikonal approximation the momentum dependence of the in-medium $\omega$ and $\eta^\prime$ width and of the inelastic meson-nucleon cross sections have been extracted. As a final result, the imaginary part of the $\omega$ and $\eta^\prime$ meson-nucleus potential as a function of the available energy has been derived. Comparing the imaginary part $W$ to the real part $V$ of the meson-nucleus potential, the $\eta^\prime$ meson is found to be a suitable candidate for the search for meson-nucleus bound states. Since $\vert W \vert << \vert V \vert$ there is a possibility for the existence of relatively narrow bound states while for the $\omega$ meson the imaginary potential is comparable to the real one and only broad structures can be expected which makes it difficult to detect them experimentally. 

An experiment to search for $\eta^\prime$ bound states via missing mass spectroscopy in the $^{12}$C(p,d) reaction in almost recoil free kinematics ~\cite{Kenta} has been performed at the Fragment Separator (FRS) at GSI and is being analysed. An alternative approach is the photoproduction of $\eta^\prime$ mesons in the  $^{12}$C($\gamma$,p) reaction, again in almost recoil-free kinematics. In the ongoing experiment at the LEPS2 facility (Spring8) \cite{Muramatsu} the missing mass spectroscopy is combined with detecting the decay of the $\eta^\prime$ mesic state in a semi-exclusive measurement. An analogous photoproduction measurement is planned at the BGO-OD setup at the ELSA accelerator in Bonn~\cite{volker}. A semi-exclusive measurement has also been considered for the Super-FRS at FAIR~\cite{Nagahiro_Kenta}. 

\section{Acknowledgements}
We thank the scientific and technical staff at ELSA and the collaborating
institutions for their important contribution to the success of the
experiment. Detailed discussions with S. Hirenzaki, H. Lenske, U. Mosel  and E. Ya. Paryev are acknowledged. This work was supported financially by the {\it
 Deutsche Forschungsgemeinschaft} within SFB/TR16 and by the {\it Schweize\-ri\-scher Nationalfonds}.

\end{document}